\documentclass[times,twocolumn]{aastex62}
\usepackage{CJK}
\usepackage{enumitem}
\usepackage{amsmath}
\usepackage{multirow}
\usepackage{cases}
\usepackage{graphicx}
\usepackage{subfigure}
\usepackage{natbib}
\usepackage{color}
\usepackage{tabularx}
\usepackage{bm}
\usepackage{threeparttable}
\usepackage{gensymb}

\newcommand{\HL}[1]{\textcolor{black}{#1}}
\newcommand{\HLb}[1]{\textcolor{black}{#1}}
\newcommand{\HLc}[1]{\textcolor{black}{#1}}
\newcommand{\HLr}[1]{#1}

\newcommand{\KMT}[2]{KMT-20{#1}-BLG-{#2}}
\newcommand{\OGLE}[2]{OGLE-20{#1}-BLG-{#2}}
\newcommand{\MOA}[2]{MOA-20{#1}-BLG-{#2}}

\newcommand{\tE}{t_{\rm E}}
\newcommand{\thetaE}{\theta_{\rm E}}
\newcommand{\thetas}{\theta_{*}}
\newcommand{\piE}{\pi_{\rm E}}

\newcommand{\Ml}{M_{\rm L}}
\newcommand{\Msun}{M_{\odot}}
\newcommand{\Mearth}{M_{\oplus}}
\newcommand{\Mjup}{M_{\rm J}}

\newcommand{\murel}{\mu_{\rm rel}}

\DeclareGraphicsExtensions{.pdf,.png,.jpg}

\shorttitle{RAMP Paper II}
\shortauthors{Yang et al.}

\begin{document}
\begin{CJK*}{UTF8}{gbsn}

\title{{\large Systematic Reanalysis of KMTNet Microlensing Events, Paper II: Two New Planets in Giant-Source Events}}

\correspondingauthor{Hongjing Yang}
\email{hongjing.yang@qq.com, yanghongjing@westlake.edu.cn}

\author[0000-0003-0626-8465]{Hongjing Yang (杨弘靖)}
\affiliation{Department of Astronomy, School of Science, Westlake University, Hangzhou, Zhejiang 310030, China}
\affiliation{Department of Astronomy, Tsinghua University, Beijing 100084, China}

\author[0000-0001-9481-7123]{Jennifer C. Yee}
\affiliation{Center for Astrophysics $|$ Harvard \& Smithsonian, 60 Garden St.,Cambridge, MA 02138, USA}

\author[0000-0002-1279-0666]{Jiyuan Zhang}
\affiliation{Department of Astronomy, Tsinghua University, Beijing 100084, China}

\author[0000-0003-0043-3925]{Chung-Uk Lee}
\affiliation{Korea Astronomy and Space Science Institute, Daejeon 34055, Republic of Korea}

\author{Dong-Jin Kim}
\affiliation{Korea Astronomy and Space Science Institute, Daejeon 34055, Republic of Korea}

\author{Ian A. Bond}
\affiliation{Institute of Natural and Mathematical Sciences, Massey University, Auckland 0745, New Zealand}

\author{Andrzej Udalski}
\affiliation{Astronomical Observatory, University of Warsaw, Al. Ujazdowskie 4, 00-478 Warszawa, Poland}

\author[0000-0002-9241-4117]{Kyu-Ha Hwang}
\affiliation{Korea Astronomy and Space Science Institute, Daejeon 34055, Republic of Korea}

\author[0000-0001-6000-3463]{Weicheng Zang}
\affiliation{Center for Astrophysics $|$ Harvard \& Smithsonian, 60 Garden St.,Cambridge, MA 02138, USA}

\author[0000-0003-4625-8595]{Qiyue Qian}
\affiliation{Department of Astronomy, Tsinghua University, Beijing 100084, China}

\author{Andrew Gould} 
\affiliation{Max-Planck-Institute for Astronomy, K\"onigstuhl 17, 69117 Heidelberg, Germany}
\affiliation{Department of Astronomy, Ohio State University, 140 W. 18th Ave., Columbus, OH 43210, USA}

\author[0000-0001-8317-2788]{Shude Mao}
\affiliation{Department of Astronomy, Tsinghua University, Beijing 100084, China}

\collaboration{(Leading Authors)}


\author[0000-0003-3316-4012]{Michael D. Albrow}
\affiliation{University of Canterbury, School of Physical and Chemical Sciences, Private Bag 4800, Christchurch 8020, New Zealand}

\author[0000-0001-6285-4528]{Sun-Ju Chung}
\affiliation{Korea Astronomy and Space Science Institute, Daejeon 34055, Republic of Korea}

\author[0000-0002-2641-9964]{Cheongho Han}
\affiliation{Department of Physics, Chungbuk National University, Cheongju 28644, Republic of Korea}

\author[0000-0002-0314-6000]{Youn Kil Jung}
\affiliation{Korea Astronomy and Space Science Institute, Daejeon 34055, Republic of Korea}
\affiliation{National University of Science and Technology (UST), Daejeon 34113, Republic of Korea}

\author[0000-0001-9823-2907]{Yoon-Hyun Ryu} 
\affiliation{Korea Astronomy and Space Science Institute, Daejeon 34055, Republic of Korea}

\author[0000-0002-4355-9838]{In-Gu Shin}
\affiliation{Center for Astrophysics $|$ Harvard \& Smithsonian, 60 Garden St.,Cambridge, MA 02138, USA}

\author[0000-0003-1525-5041]{Yossi Shvartzvald}
\affiliation{Department of Particle Physics and Astrophysics, Weizmann Institute of Science, Rehovot 7610001, Israel}


\author[0000-0002-7511-2950]{Sang-Mok Cha} 
\affiliation{Korea Astronomy and Space Science Institute, Daejeon 34055, Republic of Korea}
\affiliation{School of Space Research, Kyung Hee University, Yongin, Kyeonggi 17104, Republic of Korea} 


\author{Hyoun-Woo Kim} 
\affiliation{Korea Astronomy and Space Science Institute, Daejeon 34055, Republic of Korea}

\author[0000-0003-0562-5643]{Seung-Lee Kim} 
\affiliation{Korea Astronomy and Space Science Institute, Daejeon 34055, Republic of Korea}


\author[0009-0000-5737-0908]{Dong-Joo Lee} 
\affiliation{Korea Astronomy and Space Science Institute, Daejeon 34055, Republic of Korea}

\author[0000-0001-7594-8072]{Yongseok Lee} 
\affiliation{Korea Astronomy and Space Science Institute, Daejeon 34055, Republic of Korea}
\affiliation{School of Space Research, Kyung Hee University, Yongin, Kyeonggi 17104, Republic of Korea}

\author[0000-0002-6982-7722]{Byeong-Gon Park}
\affiliation{Korea Astronomy and Space Science Institute, Daejeon 34055, Republic of Korea}

\author[0000-0003-1435-3053]{Richard W. Pogge} 
\affiliation{Department of Astronomy, Ohio State University, 140 West 18th Ave., Columbus, OH  43210, USA}
\affiliation{Center for Cosmology and AstroParticle Physics, Ohio State University, 191 West Woodruff Ave., Columbus, OH 43210, USA}

\collaboration{(The KMTNet Collaboration)}

\author{Fumio Abe}
\affiliation{Institute for Space-Earth Environmental Research, Nagoya University, Nagoya 464-8601, Japan}

\author{Ken Bando}
\affiliation{Department of Earth and Space Science, Graduate School of Science, Osaka University, Toyonaka, Osaka 560-0043, Japan}

\author{David P.~Bennett}
\affiliation{Code 667, NASA Goddard Space Flight Center, Greenbelt, MD 20771, USA}
\affiliation{Department of Astronomy, University of Maryland, College Park, MD 20742, USA}

\author{Aparna Bhattacharya}
\affiliation{Code 667, NASA Goddard Space Flight Center, Greenbelt, MD 20771, USA}
\affiliation{Department of Astronomy, University of Maryland, College Park, MD 20742, USA}


\author{Akihiko Fukui}
\affiliation{Department of Earth and Planetary Science, Graduate School of Science, The University of Tokyo, 7-3-1 Hongo, Bunkyo-ku, Tokyo 113-0033, Japan}
\affiliation{Instituto de Astrof\'isica de Canarias, V\'ia L\'actea s/n, E-38205 La Laguna, Tenerife, Spain}

\author{Ryusei Hamada}
\affiliation{Department of Earth and Space Science, Graduate School of Science, Osaka University, Toyonaka, Osaka 560-0043, Japan}

\author{Shunya Hamada}
\affiliation{Department of Earth and Space Science, Graduate School of Science, Osaka University, Toyonaka, Osaka 560-0043, Japan}

\author{Naoto Hamasaki}
\affiliation{Department of Earth and Space Science, Graduate School of Science, Osaka University, Toyonaka, Osaka 560-0043, Japan}

\author{Yuki Hirao}
\affiliation{Institute of Astronomy, Graduate School of Science, The University of Tokyo, 2-21-1 Osawa, Mitaka, Tokyo 181-0015, Japan}

\author{Stela Ishitani Silva}
\affiliation{Department of Physics, The Catholic University of America, Washington, DC 20064, USA}
\affiliation{Code 667, NASA Goddard Space Flight Center, Greenbelt, MD 20771, USA}

\author{Yoshitaka Itow}
\affiliation{Institute for Space-Earth Environmental Research, Nagoya University, Nagoya 464-8601, Japan}

\author{Naoki Koshimoto}
\affiliation{Department of Earth and Space Science, Graduate School of Science, Osaka University, Toyonaka, Osaka 560-0043, Japan}

\author{Yutaka Matsubara}
\affiliation{Institute for Space-Earth Environmental Research, Nagoya University, Nagoya 464-8601, Japan}

\author{Shota Miyazaki}
\affiliation{Institute of Space and Astronautical Science, Japan Aerospace Exploration Agency, 3-1-1 Yoshinodai, Chuo, Sagamihara, Kanagawa 252-5210, Japan}

\author{Yasushi Muraki}
\affiliation{Institute for Space-Earth Environmental Research, Nagoya University, Nagoya 464-8601, Japan}

\author{Tutumi NAGAI}
\affiliation{Department of Earth and Space Science, Graduate School of Science, Osaka University, Toyonaka, Osaka 560-0043, Japan}

\author{Kansuke NUNOTA}
\affiliation{Department of Earth and Space Science, Graduate School of Science, Osaka University, Toyonaka, Osaka 560-0043, Japan}

\author{Greg Olmschenk}
\affiliation{Code 667, NASA Goddard Space Flight Center, Greenbelt, MD 20771, USA}

\author{Cl\'ement Ranc}
\affiliation{Sorbonne Universit\'e, CNRS, UMR 7095, Institut d'Astrophysique de Paris, 98 bis bd Arago, 75014 Paris, France}

\author{Nicholas J. Rattenbury}
\affiliation{Department of Physics, University of Auckland, Private Bag 92019, Auckland, New Zealand}

\author{Yuki Satoh}
\affiliation{Department of Earth and Space Science, Graduate School of Science, Osaka University, Toyonaka, Osaka 560-0043, Japan}

\author{Takahiro Sumi}
\affiliation{Department of Earth and Space Science, Graduate School of Science, Osaka University, Toyonaka, Osaka 560-0043, Japan}

\author{Daisuke Suzuki}
\affiliation{Department of Earth and Space Science, Graduate School of Science, Osaka University, Toyonaka, Osaka 560-0043, Japan}

\author{Sean K. Terry}
\affiliation{Code 667, NASA Goddard Space Flight Center, Greenbelt, MD 20771, USA}
\affiliation{Department of Astronomy, University of Maryland, College Park, MD 20742, USA}

\author{Paul . J. Tristram}
\affiliation{University of Canterbury Mt.\ John Observatory, P.O. Box 56, Lake Tekapo 8770, New Zealand}

\author{Aikaterini Vandorou}
\affiliation{Code 667, NASA Goddard Space Flight Center, Greenbelt, MD 20771, USA}
\affiliation{Department of Astronomy, University of Maryland, College Park, MD 20742, USA}

\author{Hibiki Yama}
\affiliation{Department of Earth and Space Science, Graduate School of Science, Osaka University, Toyonaka, Osaka 560-0043, Japan}
\collaboration{(The MOA Collaboration)}


\author{Przemek Mr\'{o}z}
\affiliation{Division of Physics, Mathematics, and Astronomy, California Institute of Technology, Pasadena, CA 91125, USA}

\author{Jan~Skowron}
\affiliation{Astronomical Observatory, University of Warsaw, Al. Ujazdowskie 4, 00-478 Warszawa, Poland}

\author{Radoslaw~Poleski}
\affiliation{Astronomical Observatory, University of Warsaw, Al. Ujazdowskie 4, 00-478 Warszawa, Poland}

\author{Micha{\l}~K.~Szyma\'{n}ski}
\affiliation{Astronomical Observatory, University of Warsaw, Al. Ujazdowskie 4, 00-478 Warszawa, Poland}

\author{Igor Soszy\'{n}ski}
\affiliation{Astronomical Observatory, University of Warsaw, Al. Ujazdowskie 4, 00-478 Warszawa, Poland}

\author{Pawe{\l} Pietrukowicz}
\affiliation{Astronomical Observatory, University of Warsaw, Al. Ujazdowskie 4, 00-478 Warszawa, Poland}

\author{Szymon Koz{\l}owski}
\affiliation{Astronomical Observatory, University of Warsaw, Al. Ujazdowskie 4, 00-478 Warszawa, Poland}

\author{Krzysztof Ulaczyk}
\affiliation{Department of Physics, University of Warwick, Gibbet Hill Road, Coventry, CV4~7AL,~UK}

\author{Krzysztof A.~Rybicki}
\affiliation{Astronomical Observatory, University of Warsaw, Al. Ujazdowskie 4, 00-478 Warszawa, Poland}

\author{Patryk Iwanek}
\affiliation{Astronomical Observatory, University of Warsaw, Al. Ujazdowskie 4, 00-478 Warszawa, Poland}

\author{Marcin Wrona}
\affiliation{Astronomical Observatory, University of Warsaw, Al. Ujazdowskie 4, 00-478 Warszawa, Poland}

\collaboration{(The OGLE Collaboration)}

\begin{abstract}
In this work, we continue to apply the updated KMTNet tender-love care (TLC) photometric pipeline to historical microlensing events. 
We apply the pipeline to a subsample of events from the KMTNet database, which we refer to as the giant source sample.
Leveraging the improved photometric data, we conduct a systematic search for anomalies within this sample.
The search successfully uncovers four new planet-like anomalies and recovers two previously known planetary signals.
After detailed analysis, two of the newly discovered anomalies are confirmed as clear planets: \KMT{19}{0578} and \KMT{21}{0736}. Their planet-to-host mass ratios are $q\sim4\times10^{-3}$ and $q\sim1\times10^{-4}$, respectively. 
Another event, \OGLE{18}{0421} (\KMT{18}{0831}), remains ambiguous. Both a stellar companion and a giant planet in the lens system could potentially explain the observed anomaly.
The anomaly signal of the last event, \MOA{22}{038} (\KMT{22}{2342}), is attributed to an extra source star.
Within this sample, our procedure doubles the number of confirmed planets, demonstrating a significant enhancement in the survey sensitivity.
\end{abstract}


\section{Introduction}\label{sec:intro}
Gravitational microlensing is the effect \HL{observed in} the light from a distant source star being lensed by a foreground stellar or planetary object. 
This phenomenon allows us to detect extrasolar planetary systems by observing the characteristic distortions in the light curves of background stars \citep{MaoPaczyski1991,GouldLoeb1992}. 
Microlensing surveys, including the Optical Gravitational Lens Experiment \citep[OGLE,][]{OGLEIII,OGLEIV}, Microlensing Observations in Astrophysics \citep[MOA,][]{Bond2001,Sumi2003_MOAobs}, Wise \citep{ShvartzvaldYossi2016_Wise}, and the Korea Microlensing Telescope Network \citep[KMTNet,][]{Kim2016_KMT}, along with the followup observations powered by them, have yielded over 200 exoplanet discoveries, proving \HL{that} microlensing \HL{is} a powerful tool for planet detection.

KMTNet, with its three 1.6~m \HL{telescopes} and 4~deg$^2$ field-of-view \HL{cameras} at the Cerro Tololo Inter-American Observatory (CTIO) in Chile (KMTC), the South African Astronomical Observatory (SAAO) in South Africa (KMTS), and the Siding Spring Observatory (SSO) in Australia (KMTA), \HL{has} been one of the most productive microlensing surveys since its \HL{commissioning in 2015}.
However, the current KMTNet survey has not reached its full potential in terms of detection sensitivity.
This limitation is primarily due to the accuracy of photometric data.

Because of the intrinsic low probability (about \HL{$10^{-6}\sim10^{-5}$ events} per monitored star per year \citep{Sumi2013_MOA2_OpticalDepth_EventRate, Mroz2019_OGLE4_OpticalDepth_EventRate}), any microlensing survey needs to observe a huge number of stars ($10^{8}\sim10^{9}$) to ensure a sufficient number of detections. 
Additionally, planetary signals are perturbations on the main light curves. The short time scale of such signals requires dense observation over time.
To meet these combined requirements, microlensing surveys must handle massive datasets with dense time sampling.
Therefore, such surveys often establish a multi-level photometric pipeline, typically consisting of an automated, general-purpose pipeline and a less automated, customized pipeline.

In KMTNet, the first levels are the DIAPL \citep{Wozniak2000_diapl} pipeline and the online pySIS \citep{pysis} pipeline. The DIAPL pipeline aims to efficiently produce the light curve of all stars within the observed field to find microlensing events. The online pySIS pipeline is then performed on those discovered events and aims to \HL{better characterize the events and detect possible anomalous signals}. Both of the above pipelines operate in real-time.
However, their inability to fully \HL{account for} the systematic errors in KMTNet images (e.g., highly irregular point spread functions, large seeing variations) limits the accuracy of the resulting data.

The second level is the Tender-Love-Care (TLC) pySIS pipeline, which aims to produce better data \HL{for detailed analysis and publication of the anomalies.} 
Following recent updates by \citet{Yang2024_pysis5_RAMP1}, this pipeline \HL{is now} more automatic and produces significantly more stable and accurate photometric data for KMTNet images.

\citet{Yang2024_pysis5_RAMP1} (hereafter Paper I) applied the updated TLC pySIS pipeline to some historical microlensing events from KMTNet \HL{and searched the new photometry for anomalous signals.}
This approach, using the improved data, successfully identified a new anomalous signal in event \MOA{19}{421}. 
Although the signal was subtle and highly degenerate, it demonstrated that this procedure can indeed enhance the detection efficiency of anomalous signals. 

In this work, we continue the idea of Paper I to systematically apply the updated TLC pySIS \HL{pipeline to} historic KMTNet events and search for new signals. As a step further than Paper I, \HL{we have two goals.}
The first is to find clear new planets, as opposed to just planet candidates as in Paper I.
The second is to evaluate the improvements of this procedure \HL{on} planet \HL{detection} efficiency. The results will determine whether it is \HL{worthwhile} to apply this procedure to all historic (and future) KMTNet events.
To achieve these two \HL{goals and limit the} workload, we decided to start with a small subsample \HL{which we estimated should have a size} of $\sim$ hundreds of events.

A straightforward choice would be the high-magnification (HM) sample (e.g., \citealt{Gould2010_mufun_stat}), as HM events are inherently more sensitive to planets. 
However, planetary signals in HM events are relatively subtle, say $\Delta A/A \sim 1\%$, where $A$ is the flux magnification due to microlensing. 
Achieving the necessary photometric precision to capture such signals often requires overcoming various \HL{sources of} systematic noise, such as the detector's non-linearity and photon-response non-uniformity.
Therefore, anomaly signals in HM events, even when identified, often require independent follow-up observations for confirmation.
\HL{Furthermore}, HM planet candidates often suffer strong degeneracies. Follow-up observations can help break \HL{those} degeneracies \citep[e.g.,][]{Jiyuan2023_kb220440}. Therefore, HM samples are better constructed through follow-up programs (\citealt{Gould2010_mufun_stat}; see also \citealt{Zang2021_KB200414_Covid19, Yang2022_kb210171_kb211689}).

In this work, we choose \HL{a} giant-source microlensing event sample.
While not as sensitive as HM events, giant-source events still have greater sensitivity to planets than the median \HL{event}.
This is because a giant source has a significant finite-source effect, which increases its probability of encountering caustics.
Moreover, planetary caustics are typically larger than central caustics \citep{Chung2005_CentralCaustic,Han2006_PlanetaryCaustic}, resulting in the majority of planetary signals arising in the low-magnification region, where they produce larger $\Delta A/A$ \citep[e.g., ``Hollywood'' events like][]{Yee2021_ob190960}. 
Additionally, giant sources are intrinsically brighter, enabling more accurate photometry.

We construct the giant source microlensing event sample using KMTNet survey data from the 2016 to 2022 seasons\footnote{\HL{excluding 2020, which was impacted by the COVID pandemic}}. The subsample contains 352 events. 
We apply the updated TLC pySIS pipeline to all events in this sample to yield accurate light curves. 
We subsequently developed an anomaly search algorithm to identify any anomalous signals within these events, successfully uncovering four new planet-like anomalies. 
The structure of this paper is as follows. Section \ref{sec:pipeline} presents some new updates of the photometric pipeline. 
Section \ref{sec:sample} describes the sample selection.
After applying the pipeline, we describe the anomaly search in Section \ref{sec:af}. 
In Section \ref{sec:analysis}, we present the modeling of the newly discovered signals.
Finally, Section \ref{sec:summary} summarizes our findings based on this sample.
A detailed analysis of the sample's sensitivity and statistical properties will be presented in future work.

\section{Photometry Pipeline}\label{sec:pipeline}
Building upon the work presented in Paper I \citep[see Section 2 in][for the overall procedures and details of the photometry pipeline]{Yang2024_pysis5_RAMP1}, we have further refined the photometric pipeline to enhance its self-consistency and facilitate its application to a larger sample.

First, for the photometric algorithm, we now consider (linear) inter-pixel correlations. 
In Paper I, we develop a ``blur'' algorithm to handle \HL{images with very small full-width half-maximum (FWHM).} However, this process introduced undesirable pixel-to-pixel correlations, potentially leading to underestimates of the photometric errors.
To address this issue, we now consider the linear correlation between a pixel and its left and upper neighbors.
In the non-correlated linear optimizations, we minimize
\begin{equation}
    \chi^2_{\rm non-corr.} = \sum_{i,j}{\frac{\epsilon_{i,j}^2}{\sigma_{i,j}^2}}, \quad
    \epsilon_{i,j} = \mathcal{D}_{i,j} - (b+f\mathcal{P}_{i,j}),
\end{equation}
to obtain the difference flux $f$ and background $b$.
Where $\mathcal{D}_{i,j}$, $\mathcal{P}_{i,j}$ and $\sigma_{i,j}$ are the $(i,j)$-th pixel value of the difference image, the pixelated PSF model, and the uncertainty image, respectively. 
This process assumes independent and Gaussian-distributed $\epsilon_{i,j}$ values.
Now, we consider a correlated error,
\begin{equation}
    \epsilon_{i,j} = c_1\epsilon_{i-1,j} + c_2\epsilon_{i,j-1} + \nu_{i,j},
    \label{eq:linfit_1}
\end{equation}
where $c_1$ and $c_2$ are the constant correlation coefficients, and $\nu_{i,j}$ is the de-correlated error which should be independent and Gaussian-distributed.
Then, the value to be minimized becomes
\begin{equation}
    \chi^2_{\rm corr.} = \sum_{i,j}{\frac{\nu_{i,j}^2}{\sigma_{i,j}^2}}.
\end{equation}
This minimization corresponds to a linear fit applied to the de-correlated difference image ($\mathcal{D}'_{i,j}$) and the de-correlated PSF model ($\mathcal{P}'_{i,j}$) 
\begin{equation}
    \mathcal{D}'_{i,j} = b+f\mathcal{P}'_{i,j} +\nu_{i,j},
\end{equation}
where
\begin{equation}
    \mathcal{D}'_{i,j} = \mathcal{D}'_{i,j} - c_1\mathcal{D}'_{i-1,j} - c_2\mathcal{D}'_{i,j-1}
\end{equation}
and
\begin{equation}
    \mathcal{P}'_{i,j} = \mathcal{P}'_{i,j} - c_1\mathcal{P}'_{i-1,j} - c_2\mathcal{P}'_{i,j-1}.
\end{equation}
The constant correlation coefficients ($c_1$ and $c_2$) are determined by a separate linear fit to $\epsilon_{i,j}$.
In practice, we begin by assuming $c_1=c_2=0$, calculate $(f,b)$, and then iteratively use these results to estimate $(c_1,c_2)$. To prevent unphysical values, we limit the correlation coefficients to be positive and less than 0.4. This constraint is consistent with the ``blur'' process.

Second, we updated the definition of the photometry goodness indicator to be
\begin{equation}
    \sigma_{\rm res} = \frac{1}{N_{\rm phot}}\sum_{i,j}{\frac{|\mathcal{\epsilon}_{i,j}|}{\sigma_{i,j}^2 + \sigma^2_{\rm RON}}} 
\end{equation}
(Notation consistent with Paper I, Section 2.1).
This new definition incorporates the noise term for a more general representation.
Previously, the threshold for ``bad'' data points varied across events. The revised indicator allows for a universal threshold of $\log \sigma_{\rm res}<0.5$ which works for most events.

Lastly, we construct a list of the CCD defects for all KMTNet cameras so that those pixels can be automatically masked during image subtractions.


Following these improvements, the pipeline (hereafter referred to as ``auto-TLC'') can now be efficiently and automatically applied to a large sample, minimizing the need for manual inspection.

\section{Sample Selection}\label{sec:sample}
Giant stars, with their larger radii, offer advantages in microlensing planet detection. 
The finite source size $\rho$ is defined by
\begin{equation}
    \rho \equiv \frac{\thetas}{\thetaE},
\end{equation}
where $\thetas$ and $\thetaE$ are the angular radius of the source star and the Einstein ring, respectively. 
The typical Einstein radius for Galactic microlensing events ranges from $\sim0.1$ to $1~{\rm mas}$. 
The source sizes of main-sequence stars and red giants in the Galactic bulge are typically $\sim 0.4-0.8~{\rm \mu as}$ and $\sim3-15~{\rm \mu as}$, corresponding to $\rho$ values of $\sim 10^{-3}$ and $\sim 10^{-2}$, respectively.
When the planet-host separation is large or planet mass is small, the size of the planetary caustic can be less than $10^{-2}\thetaE$ \citep{Han2006_PlanetaryCaustic}. 
In these cases, the probability of a caustic crossing is dominated by $\rho$ rather than the caustic size itself.
Therefore, the giant sources are relatively more sensitive to these planets.

Before outlining the specific criteria for our giant source sample, we list a few key expectations.
First, the sample should mostly consist of giant sources, so that the advantages of giant sources (bright, sensitive to planets) can be exploited. 
A small fraction of non-giant source events (e.g., a giant star acting as a blending source) is acceptable, provided the sample remains unbiased in terms of planet detections.
Second, the sample size should be of \HL{order} a few hundred events. This ensures that the photometry and anomaly search can be completed within a reasonable timescale ($\sim$ months) given current computational resources and manpower. 
As we described in Section \ref{sec:intro}, the purpose of this starting subsample is, on the one hand, to discover clear previously missed planets, and on the other hand, to systematically confirm whether further research on a larger sample is \HL{worthwhile}.

We describe the detailed sample selection criteria below. 
The selection uses the information from the KMTNet website\footnote{\url{https://kmtnet.kasi.re.kr/~ulens/}}. 
We focused on events discovered between 2016 and 2022 seasons for which seasons the EventFinder \citep{Kim2018a_EventFinder} has been completed. 
Events from the 2020 season are excluded. The 2020 event search is incomplete because both CTIO and SAAO were shut down during the COVID-19 pandemic. 
For the remaining events, the criteria are as below:
\begin{enumerate}[wide,label=(\arabic*).,itemindent=!] 
    \item $I_{\rm cat}-A_I<16$,
    \item $I_{\rm s}-A_I<16$,
    \item $I_{\rm baseline}<17$, 
    \item Located in fields with \HL{combined} cadence $\leq 15~{\rm min}$.
\end{enumerate}
The first criterion is designed to select giant stars \citep{Kim2021_FFP_KB192073}, \HL{down to} approximately 1.5 magnitudes below the red clump center. 
The second criterion ensures the source indeed corresponds to the cataloged giant star.
The third and fourth criteria do not hold specific physical meaning but only limit the observed signal-to-noise ratio and sample size.

As previously mentioned, giant source events are sensitive to \HL{wide-orbit} planets. However, the standard KMTNet TLC pipeline is typically only applied to the event season itself, as most events are shorter than one year.
In this work, \HL{to allow} searching \HL{for} distant signals \HL{far} from the peak, \HL{images from multiple years} are included to produce the photometric data. 
Because the probability of finding a $s>15$ signal drops to almost \HL{nothing}, we include \HL{only} images within $t_0 \pm 15 t_{\rm E}$ for each event. If a portion of a season overlaps with the above time range, the entire season is used.

Following these criteria, a total of 352 events were selected. 
We then employ the auto-TLC pipeline for a systematic photometric re-reduction of these events.
The majority of images of the KMTNet survey were taken in the $I$ band, and about 9\% were taken in the $V$ band. The systematic re-reduction includes all $I$-band images and KMTC $V-$band images from the selected seasons for color measurements.

The auto-TLC pipeline encountered failures in 12 events. 
These events were all located near field boundaries, where the pipeline failed in automatic reference image selection. The failure rate ($12/352\approx3.4\%$) is consistent with \HL{the frequency of stars} near the boundary of a field $300/9000\approx3.3.\%$. 
Because the planet occurrence rate should not depend on the distance to the image boundary, we do not \HL{try to recover} these events. 
Consequently, our subsequent anomaly signal search will focus on the remaining 340 events.

\section{Anomaly Search}\label{sec:af}
\subsection{The Anomaly Search Algorithm}
    \citet{Zang2021_ob191053_AnomalyFinder1} highlighted the need for a search algorithm to comprehensively find the anomalies, particularly for a survey like KMTNet. 
    Following this idea, here we develop a new algorithm to automatically detect candidate signals.
    
    The first step for searching anomalies in a light curve, is to remove ``normal'' signals. 
    In our case, the ``normal'' signal is the standard point-source point-lens (PSPL) microlensing light curve \citep{Paczynski1986}. Three parameters are required to describe the magnification as a function of time, $A(t)$. They are $u_0$, the impact parameter of the lens-source relative trajectory (in units of the Einstein radius), $\tE$, the microlensing timescale, and $t_0$, the closest time of the lens and source. The magnification is then
    \begin{equation}
        A(t) = \frac{u(t)^2+2}{u(t)\sqrt{u(t)^2+4}}, \quad 
        u(t) = \sqrt{u_0^2 + \left(\frac{t-t_0}{\tE}\right)^2}.
    \end{equation}
    Two flux parameters, $f_{\rm S}$ and $f_{\rm B}$ for each site and field are needed to represent the magnified and unmagnified fluxes.
    For each event, we fit the entire light curve with such a PSPL model and extract the residuals. Subsequent anomaly searches are mainly based on the residuals.
    
    The existing KMTNet AnomalyFinder \citep{Zang2021_ob191053_AnomalyFinder1,Zang2022_AnomalyFinder4_2019prime} employs a robust strategy for searching for signals in the online data, which are often noisy and contain many outliers. This approach involves fitting a series of PSPL models to the residuals.
    However, for the high-quality auto-TLC data used in this work, the photometric algorithms effectively identify and exclude most outliers, significantly reducing the false positives. Therefore, to improve the efficiency, we developed a new anomaly search algorithm that \HL{does not require a model for the anomaly}. 
    
    Despite the improved stability of the auto-TLC data compared to online data, false positives can still occur due to reasons like poor seeing conditions, high sky background, defects in the image, or intrinsic variability in the source or the nearby blended stars.
    Therefore, we design our algorithms accordingly to reduce false positives caused by these reasons.
    
\begin{figure*}
    \centering
    \includegraphics[width=1.7\columnwidth]{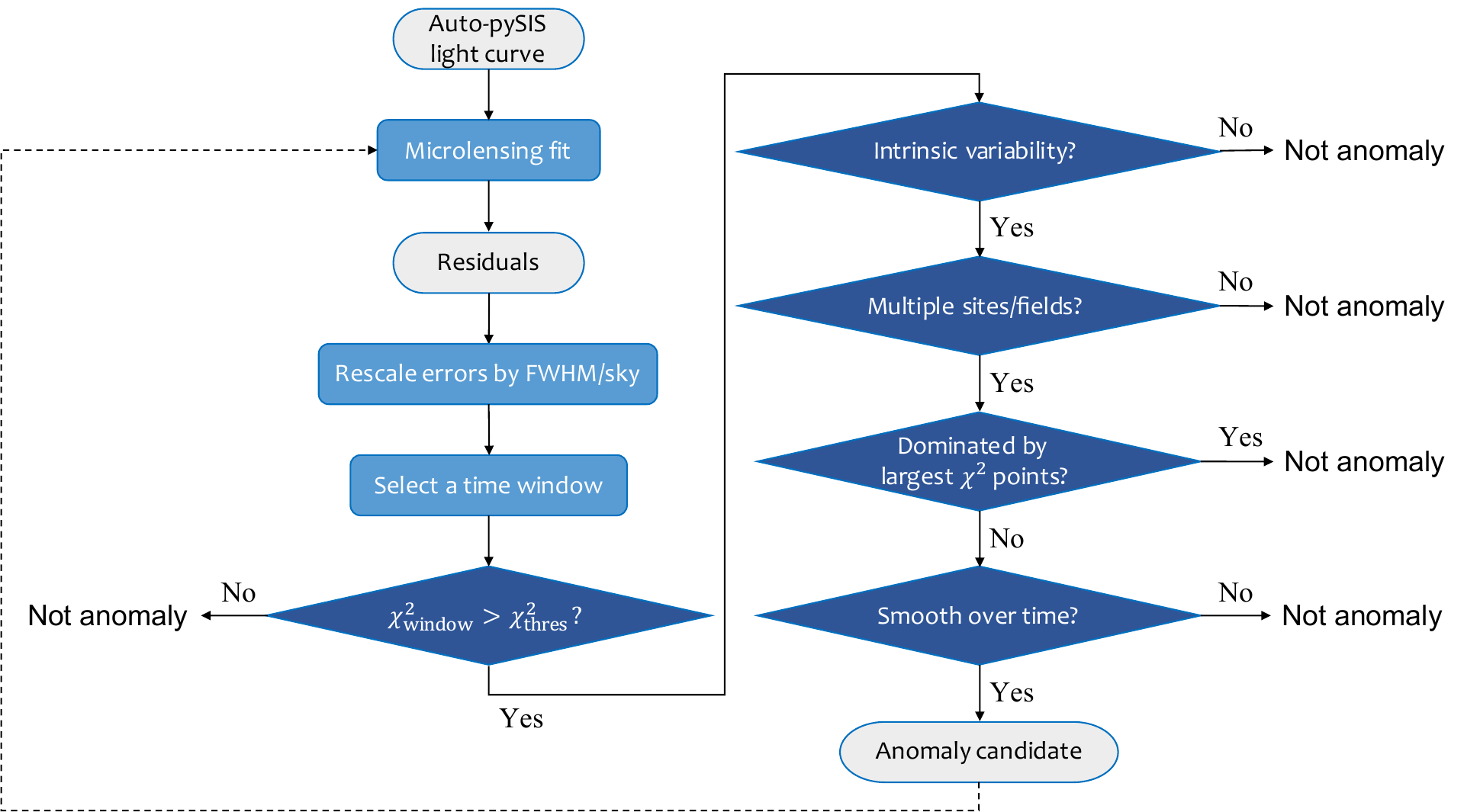}
    \caption{Flow chart of the anomaly search.}
    \label{fig:flow_chart}
\end{figure*}

    The flow chart outlining our anomaly search procedure is presented in Fig. \ref{fig:flow_chart}. Here we first briefly list all the main steps and then follow this with a detailed explanation of each step.
    
    \begin{enumerate}[wide,label=(\alph*).,itemindent=!]
        \item Fit the light curve with microlensing models and extract the residuals.
        \item Rescale the error bars of the data in FWHM bins for each site/field.
        \item Define a time window for searching for anomalies.
        \item In this time window, calculate the cumulative $\chi^2$ and compare it to a threshold.
        \item Check if the signal is significant compared to the baseline variability.
        \item Check if the window contains data from multiple fields/sites. 
        \item Check if the signal is dominated by a small fraction of largest $\chi^2$ points.
        \item Check if the signal comes from smooth deviations (consistent over time) rather than scattered points.
        \item If a signal passes all these criteria, it is labeled as an anomaly candidate.
    \end{enumerate}
    
    When fitting the light curve in (a), we first use the static point-source point-lens (PSPL) model. We fit the light curve and extract the residuals \HLr{$\Delta F$ (observed flux $-$ model) and their corresponding uncertainties $\sigma_F$} in flux space.
    Initially, any deviations from the static PSPL model are considered as potential anomaly candidates.
    
    The purpose of step (b) is to reduce the false positives caused by seeing variations. Although the auto-TLC data are much more stable than the online data, correlations between flux and seeing still exist. 
    Therefore, we rescale the error bars of the data points in a series of FWHM bins for each site/field. \HLr{To allow each FWHM bin to have sufficient points, the bins are taken from the minimum to the maximum value with a bin width of $0.2^{\prime\prime}$.}
    We calculate $\chi^2\big/N_{\rm pt}$ (where $N_{\rm pt}$ is the number of data points) in these seeing bins, and use it to determine the scaling factor $k$,
    \begin{equation}
        \sigma_{F,i}^\prime = k \sigma_{F,i},
    \end{equation}
    that adjusts the errors to achieve
    \begin{equation}
        \chi^2 = \sum_i\left(\frac{\Delta F_i}{\sigma_{F,i}^\prime}\right)^2 = 1,
    \end{equation}
    \HLr{where $\Delta F_{i}$, $\sigma_{F,i}$, and $\sigma_{F,i}^{\prime}$ are the residual, and the uncertainties of each measurement before and after the rescaling.}
    Because the light curve covers at least 2 yr for these events, the $\chi^2$ here is dominated by the unmagnified baseline data. Planetary signals, even if they exist, are only a small fraction of the whole light curve. Therefore, the false negatives caused by this step are \HL{negligible}, but many false positives are reduced. Nevertheless, we limit $k$ to be $1.0\leq k\leq3.0$ \HL{to prevent} making false negatives. 
    This step also reduces the influence of intrinsic source variability on false positives.
    
    In (c), to find signals as short as $\sim 0.2~$d, we sample 16 logarithmic uniform values in  $0.02~{\rm d} - 2~{\rm d}$ and 20 logarithmic uniform values in $2~{\rm d} - 2000~{\rm d}$ as the \HL{duration} of the time window. 
    For each \HL{window duration}, we set the time window to start from the first data point and shift it with a step size of $1/10$ of the window length until the data ends. On average, this approach divides the data for each year into $\sim$600,000 windows. 
    
    Then in (d), we first skip windows containing $\leq3$ data points because they are unreliable for the anomaly detection. On average, $75\%$ of windows are skipped \HL{due to this criterion}. For the windows with sufficient data points, $N_{\rm window}$, we calculate the cumulative $\chi^2$ in the window using
    \begin{equation}
        \chi^2_{\rm window} = \sum_{i\in {\rm window}}{\left(\frac{\Delta F_i}{\sigma_{F,i}}\right)^2}.
    \end{equation}
    where the subscript $i$ represents the index of the data point.
    Assuming the residuals follow an independent Gaussian distribution (no signal present),  the $\chi^2_{\rm window}$ should follow the \HL{$\chi^2$} distribution. 
    We select the threshold $\chi^2_{\rm thres}(N_{\rm window})$ as a function of $N_{\rm window}$, such that random variables from a \HL{$\chi^2$} distribution have a probability less than $6.33\times10^{-5}$ (corresponding to a $>4\sigma$ probability for a Gaussian distribution) of exceeding the threshold. 
    Larger $\chi^2_{\rm window}$ are considered anomalous and indicative of potential anomaly signals. 
    \HLr{Additionally, a minimum threshold of $\chi^2_{\rm thres,min}(N_{\rm window})=N_{\rm window}+80$ is set to ensure the signal is significant enough compared to unrecognized non-Gaussian systematic errors. Similar minimum thresholds are widely used in previous microlensing planet detections \citep[e.g., ][]{Suzuki2016,Poleski2021_WideOrbitStatistic,Zang2021_ob191053_AnomalyFinder1}, where the exact value varies regarding the features of the dataset.}
    Only windows with $\chi^2_{\rm window}>\chi^2_{\rm thres}$ proceed to the next steps.
    
    In (e), we check the source variability. 
    Because giant sources often have intrinsic variability, we ensure reported anomalies are not dominated by these variations.
    The baseline variability is defined as the root-mean-square of $\Delta F/\sqrt{F}$, calculated as
    \begin{equation}
        {\rm RMS} = \sqrt{\frac{1}{N_{\rm data}}\sum_{i}^{N_{\rm data}}{\frac{\Delta F_i^2}{F_i}}},
    \end{equation}
    where $N_{\rm data}$ is the total number of data points \HLr{and $F_i$ is the original measured flux}. 
    The reason why we adopt RMS of $\Delta F/\sqrt{F}$ instead of $\Delta F$ is that the magnified points tend to have larger $\Delta F$s. 
    We therefore normalize $\Delta F$ by $\sqrt{F}$ to account for Poisson errors during microlensing events.
    Similarly, the RMS for a window is calculated as
        \begin{equation}
            {\rm RMS_{window}} = \sqrt{\frac{1}{N_{\rm window}}\sum_{i\in {\rm window}}{\frac{\Delta F_i^2}{F_i}}}.
        \end{equation}
    Assuming the residuals $\Delta F/\sigma_F$ follow independent Gaussian distributions, then $(\sqrt{N_{\rm window}}{\rm RMS_{window}}/{\rm RMS})$ should follow a \HL{$\chi$} distribution. We adopt a threshold corresponding to $<4.55\%$ probability ($>2\sigma$ probability for a Gaussian distribution). If the variation in a window exceeds this threshold, the window proceeds to the next step.
    
    In (f), we simply check if the window contains data from more than one site or field so that the signals can be cross-verified. If true, it is passed to the next step.
    
    Then, in (g), we check if the signal is dominated by a few outliers. In many cases, the light curve can occasionally contain outliers. 
    Real planetary signals in giant source events typically have a timescale of $t_{\rm anom}\sim t_{*}=\rho\tE$ and thus cannot be extremely short \HL{or consist of only a few data points}.
    Therefore, here we exclude the $5\%$ (rounded down to an integer value, with a minimum of 1) highest $\chi^2$ points and re-evaluate the remaining points against the criterion in step (d). Note that because the number of points changes, the threshold also changes.
    This step also helps exclude cases where a very large window contains a very short anomaly, improving the efficiency for human reviewers.
    
    Finally, in (h), we check if the ``signal'' is consistent over time, i.e., not just caused by randomly scattered points. \HL{Random} signals are not smooth over time and are not considered candidate signals. 
    We fit the residuals in the time window with a polynomial function and extract the corresponding $\chi^2_{\rm poly}$. Based on testing results, we empirically choose the polynomial order to be $\lfloor N_{\rm window}/6\rfloor$, with minimum order $2$ and maximum order $8$. We then check if $\chi^2_{\rm poly}$ satisfies
    \begin{equation}
        \frac{\chi^2_{\rm poly}-N_{\rm window}}{\chi^2_{\rm window}-N_{\rm window}}<0.3,
    \end{equation}
    i.e., if the polynomial model can explain $\geq70\%$ signal, $\Delta\chi_{\rm window}^2=\chi^2_{\rm window}-N_{\rm window}$. 
    If a signal is caused by randomly scattered points, then $\chi^2_{\rm poly}$ will be comparable to $\chi^2_{\rm window}$, because a smooth polynomial function cannot reproduce the randomly scattered ``signals''. On the other hand, if the ``signal'' is smooth over time, one can find $\chi^2_{\rm poly}$ significantly smaller than $\chi^2_{\rm window}$.
    Signals that pass all the above criteria are classified as anomaly candidates and await further inspection and/or modelling.
    
    Finally, for a light curve containing possible signals, the above search often returns many overlapping windows, which can be troublesome for human reviewers. Therefore, we group these reported windows as follows: If a smaller window is fully included in a larger window, we keep the one with larger $\Delta\chi_{\rm window}^2$ and discard the other. The process starts from the smallest window, and continues until all windows are not fully included in each other. The final windows are reported to the reviewer.
    
    Now, the anomaly search algorithm has been built. It is named easyAnomalyFinder or \texttt{eAF} for short. 
    \HLr{Because it avoids detailed model fitting, the typical time cost per event is only $\sim1-5$ seconds on a single CPU core, which will eventually speed up the sample sensitivity calculation. Relevant results will be presented in future work.}
    \HLr{We note that the \texttt{eAF} algorithms can potentially have a broader usage, such as searching for various types of time-domain signals. We managed to make the code easier to customize. The reader can find the public code on \href{https://github.com/hongjingy/easyAnomalyFinder}{GitHub}\footnote{\url{https://github.com/hongjingy/easyAnomalyFinder}}. }
    
    We then apply \texttt{eAF} to the auto-TLC data of the 340 giant source events. 
    We review the reported signals, if they are false positives, we clean the data accordingly and search for the signals again. If the signals are likely true, they are modeled, and the residuals from the new model are fed back into \texttt{eAF} until no signals are reported. 
    The search results are described in the next section.

\subsection{Apply easyAnomalyFinder to Giant Source events}\label{sec:eAF:apply}
    We apply \texttt{eAF} to the 340 events with new photometric data. Here we present the results.
    We first classify the events into two categories: events with no signal (or a signal caused by systematic errors or intrinsic variation) and events with candidate signals. 
    A total of 59 events have candidate signals, while 281 have no signal.
    
\begin{figure*}
    \centering
    \includegraphics[width=0.9\columnwidth]{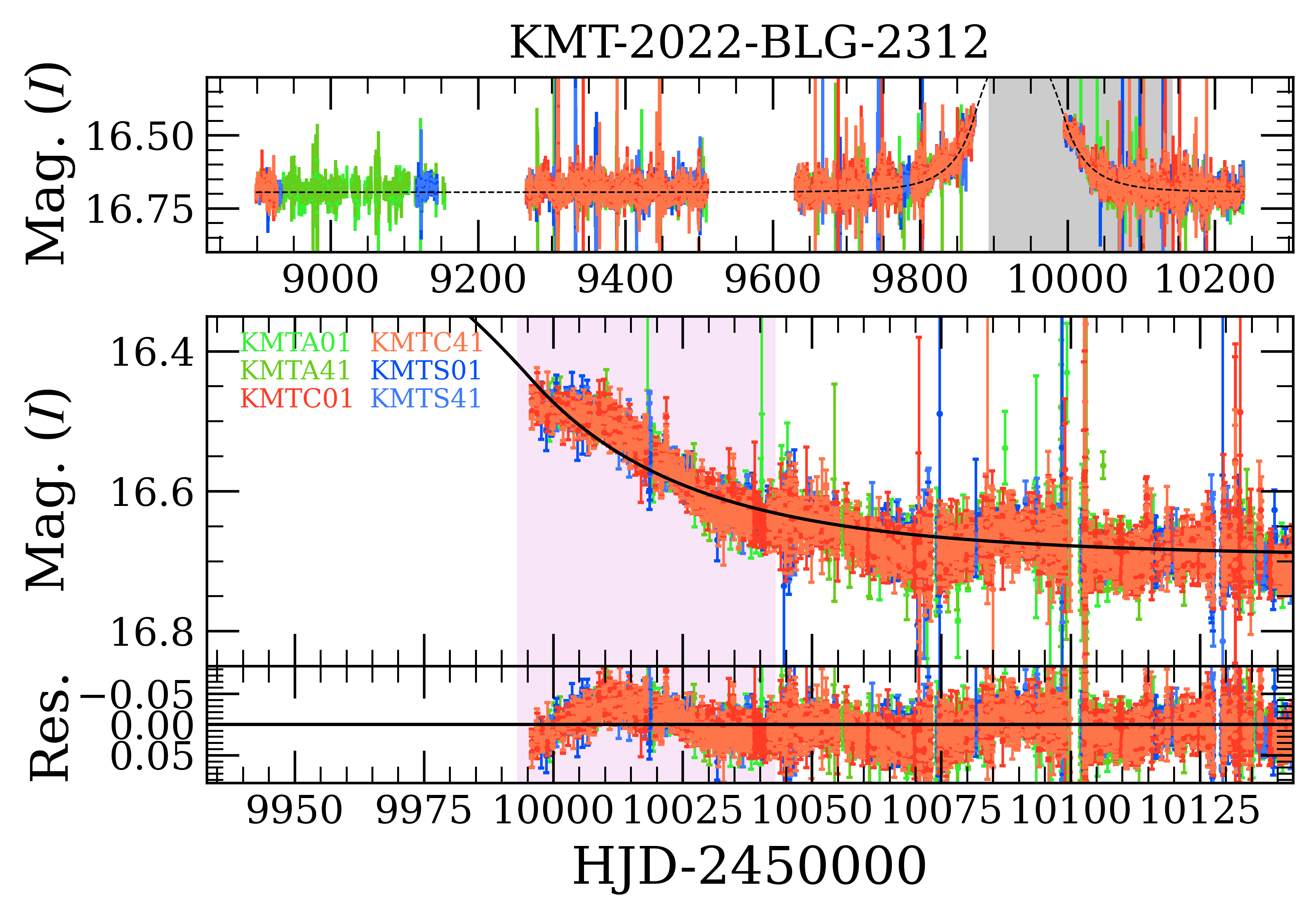}
    \includegraphics[width=0.9\columnwidth]{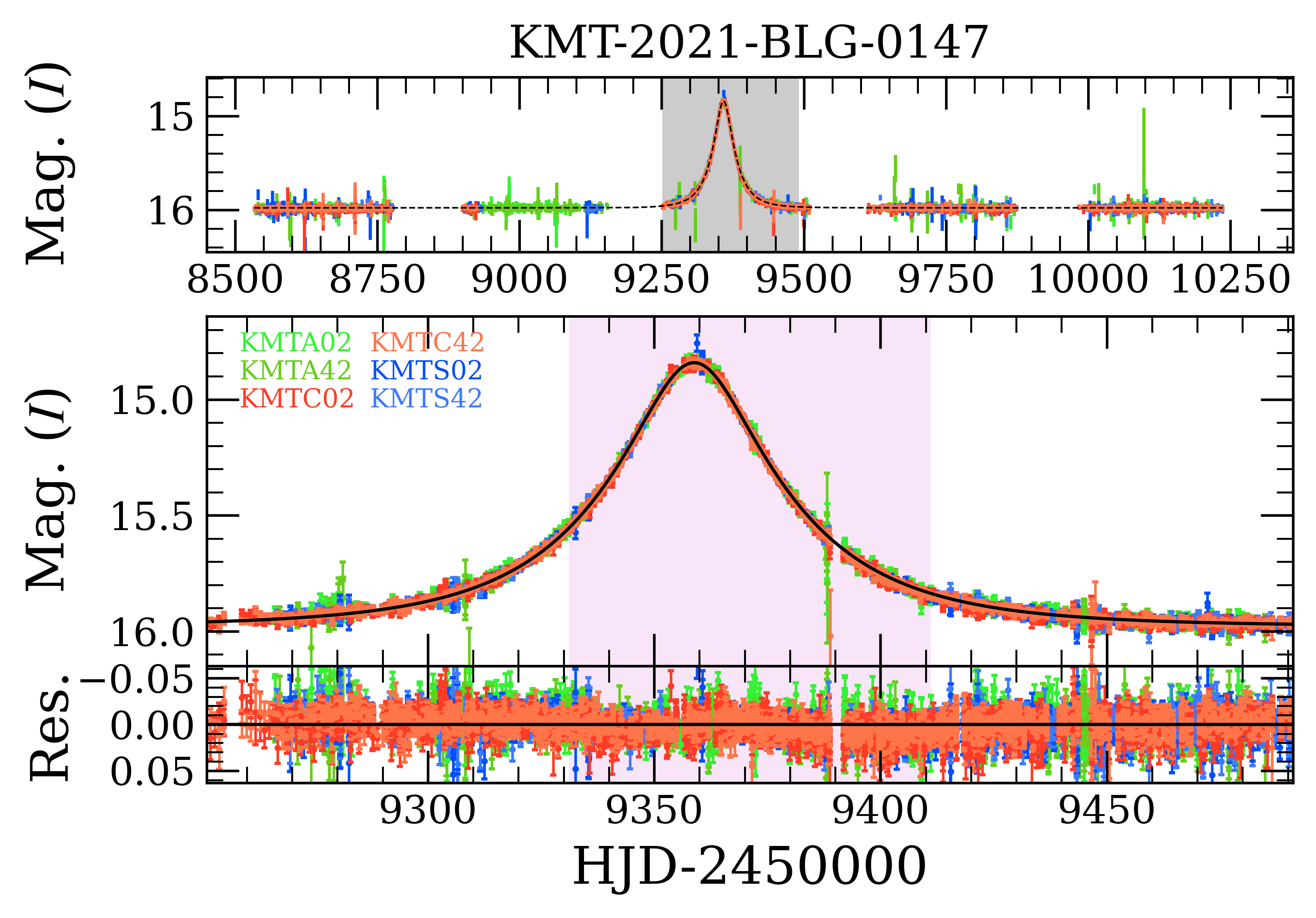}
    \\ \vspace{10pt}
    \includegraphics[width=0.9\columnwidth]{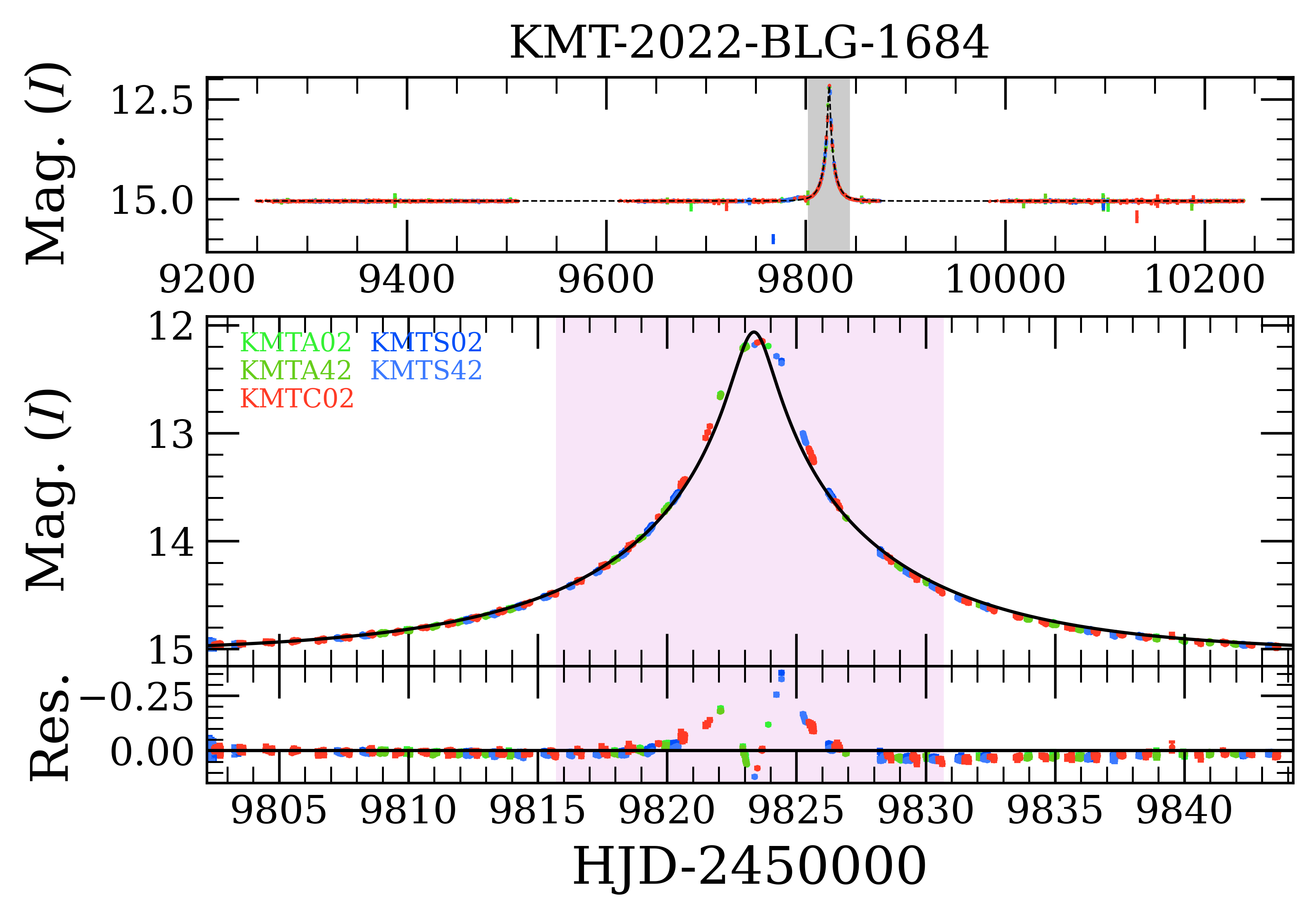}
    \includegraphics[width=0.9\columnwidth]{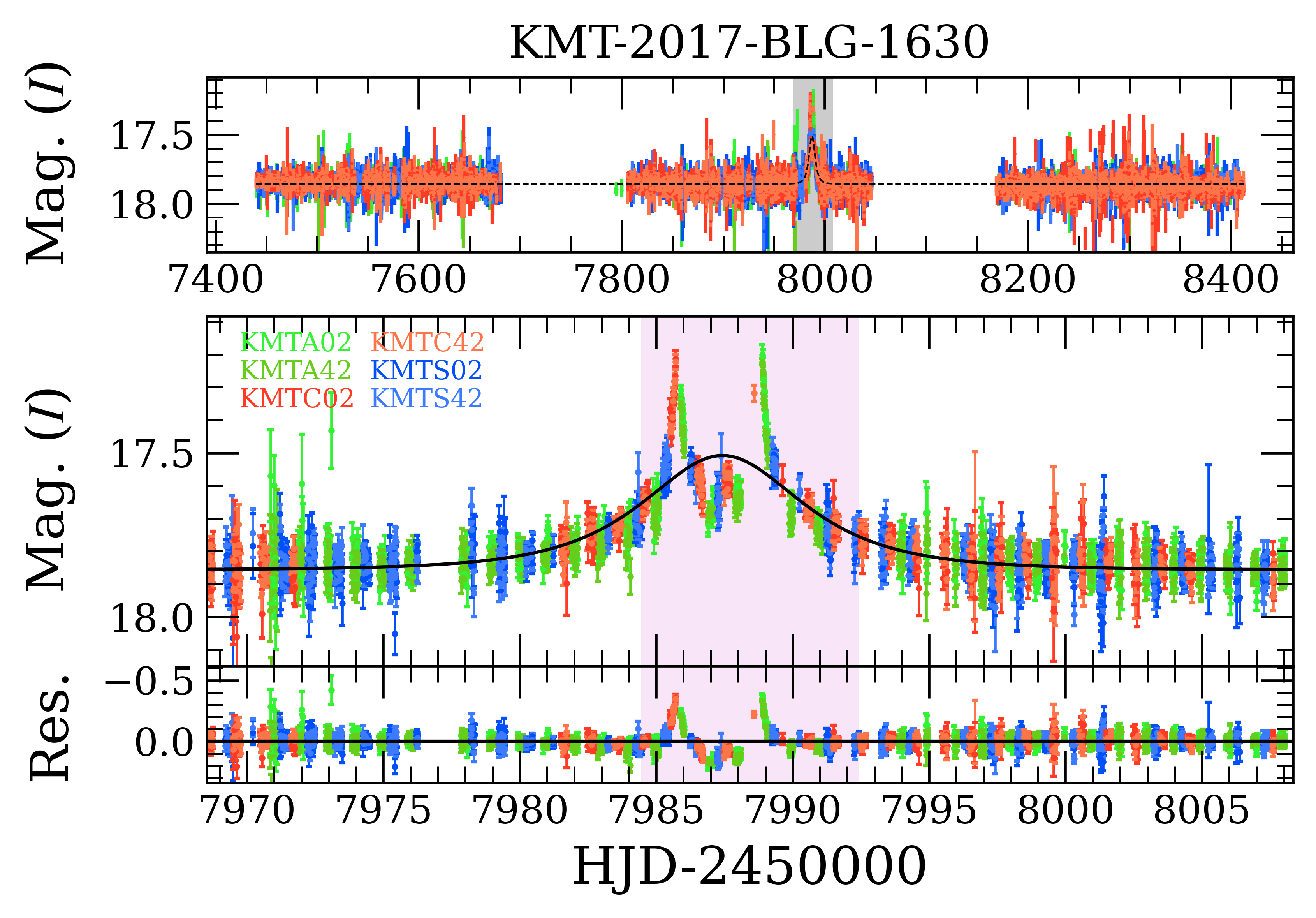}
    \caption{Light curves and reported signals of some example candidate \HL{anomalies}. In each subfigure, the upper panel is the full light curve \HL{of the event} and the lower is the zoom-in plot of the candidate signal region. The shadowed region marks the reported time window that has the candidate signal. The colors represent data from different sites and fields. The PSPL models are shown as black curves.}
    \label{fig:anomlc_noplanet}
\end{figure*}

    In the no-signal group, 235 do not report any signals or the ``signal'' is due to outliers that can be easily eliminated by cleaning the data. 
    Seven events are not real microlensing events but false positives of the KMTNet AlertFinder \citep{Kim2018_KMTalgorithm} or EventFinder \citep{Kim2018a_EventFinder}. 
    Thirty-four events exhibited very significant intrinsic variability, making it unlikely any real signals can be identified within them. An example is shown in the upper left panel of Fig. \ref{fig:anomlc_noplanet}. 
    The other 5 events have significant systematic errors that could not be cleaned.
    
    Among the 59 events with candidate signals, we cross-matched them with known planetary and binary catalogs and further categorized them as: known planetary events, known stellar binary events, and new anomalous events.
    All the known planets \HL{(2)} and known binaries \HL{(16)} in this sample are successfully recovered. The two known planets are \OGLE{18}{0383}/\KMT{18}{0900} \citep{Wang2022_AnomalyFinder3_kb180900_ob180383} and \OGLE{16}{0007}$/$\KMT{16}{1991} \citep{Zang2024_science_KMTMF}. 
    The remaining 41 events are considered new anomalous events that require further investigation.

    \HLr{Sixteen} of these events display asymmetric signals like \KMT{21}{0147} (shown in Fig. \ref{fig:anomlc_noplanet}, upper right panel). These features could be caused by the microlensing parallax effect \citep{Gould1992,Gould2000_Formalism,Gouldpies2004}. \HL{This effect} is a result of Earth's orbital motion, which provides acceleration to the source-lens-Earth relative motion. 
    Here we include the parallax effect in the modelling of these events, obtaining new residuals. 
    No further signals were found using \texttt{eAF} on these new residuals.
    Therefore, we classify these events as possible parallax events.
    Note that some of these signals could be due to the ``xallarap'' effect, i.e., the source star's orbital motion \HL{around a companion} \citep{Griest1992_Effect, Han_Gould_1997_1L2Sxarallap}. 
    In many cases, the parallax and xallarap effects are degenerate. 
    However, this work focuses on planetary candidates, so we do not investigate these signals in detail for now.
    
    In the remaining \HLr{25} events, 4 of them have ``flat-top'' features. An example is shown in the lower left panel of Fig. \ref{fig:anomlc_noplanet}. 
    These features are likely caused by the finite-source effect when $\rho>u_0$. 
    However, the extended source could also cross the cental caustic produced by a planet in the lens. 
    Therefore, we update the models to finite-source point-lens (FSPL) models and re-evaluate the residuals. 
    One event (\KMT{22}{0330}) shows no clear residuals after FSPL fitting, while three still have signals. We then search for a finite-source binary-lens (FS2L) model for these three events (methods are the same as will be described in Section \ref{sec:ana:pre}). Finally, we find all of them are clear stellar binary events with secondary-to-primary mass ratio $q>0.1$. 
    
    We fit for single-source binary-lens (1S2L) models for \HL{the remaining} \HLr{21} candidate events. We find \HLr{17} of them are clear stellar binary events ($\log q\geq-1.5$) like \KMT{17}{1630} (see lower right panel of Fig. \ref{fig:anomlc_noplanet}). 
    
\begin{figure*}
    \centering
    \includegraphics[width=0.9\columnwidth]{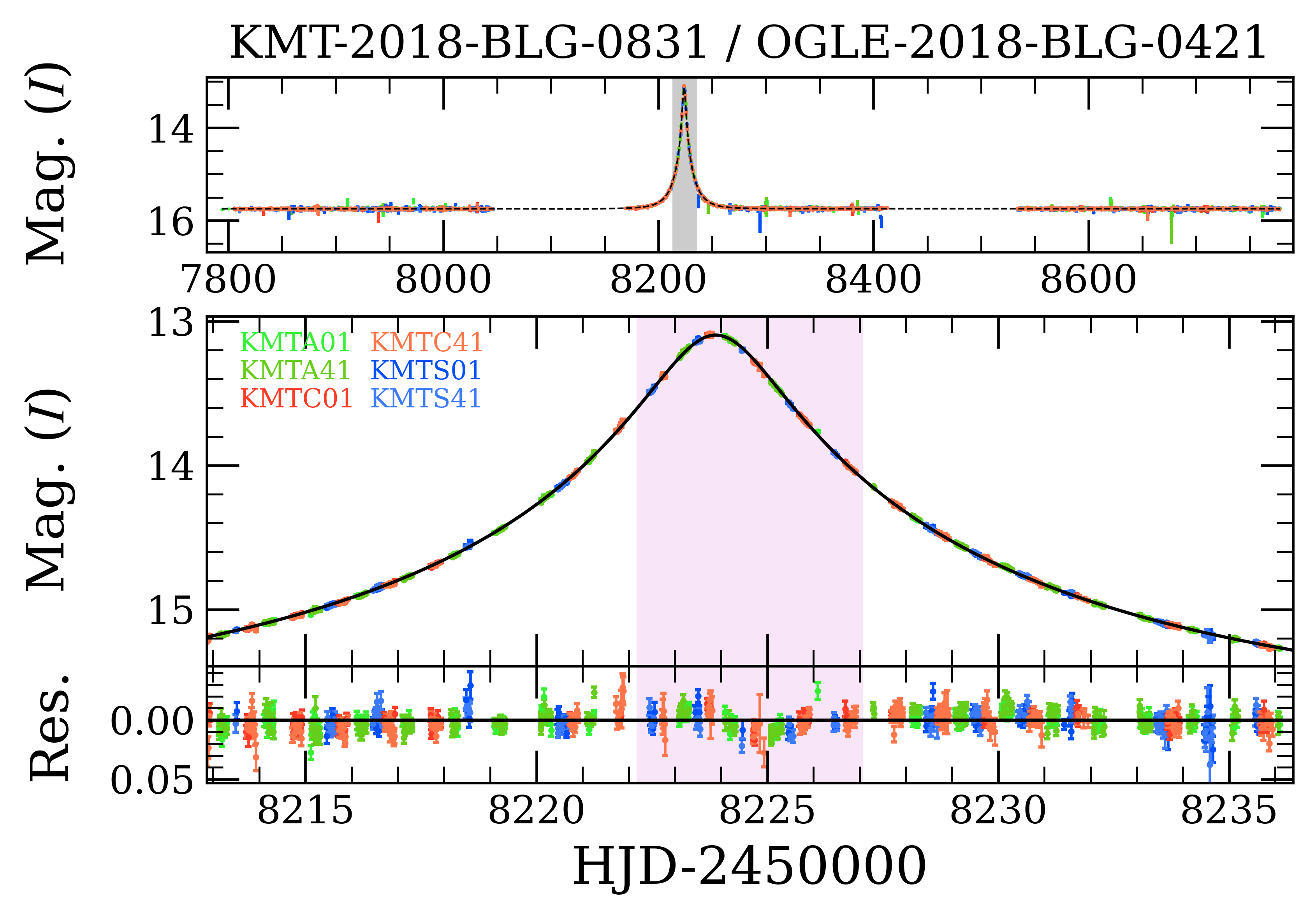}
    \includegraphics[width=0.9\columnwidth]{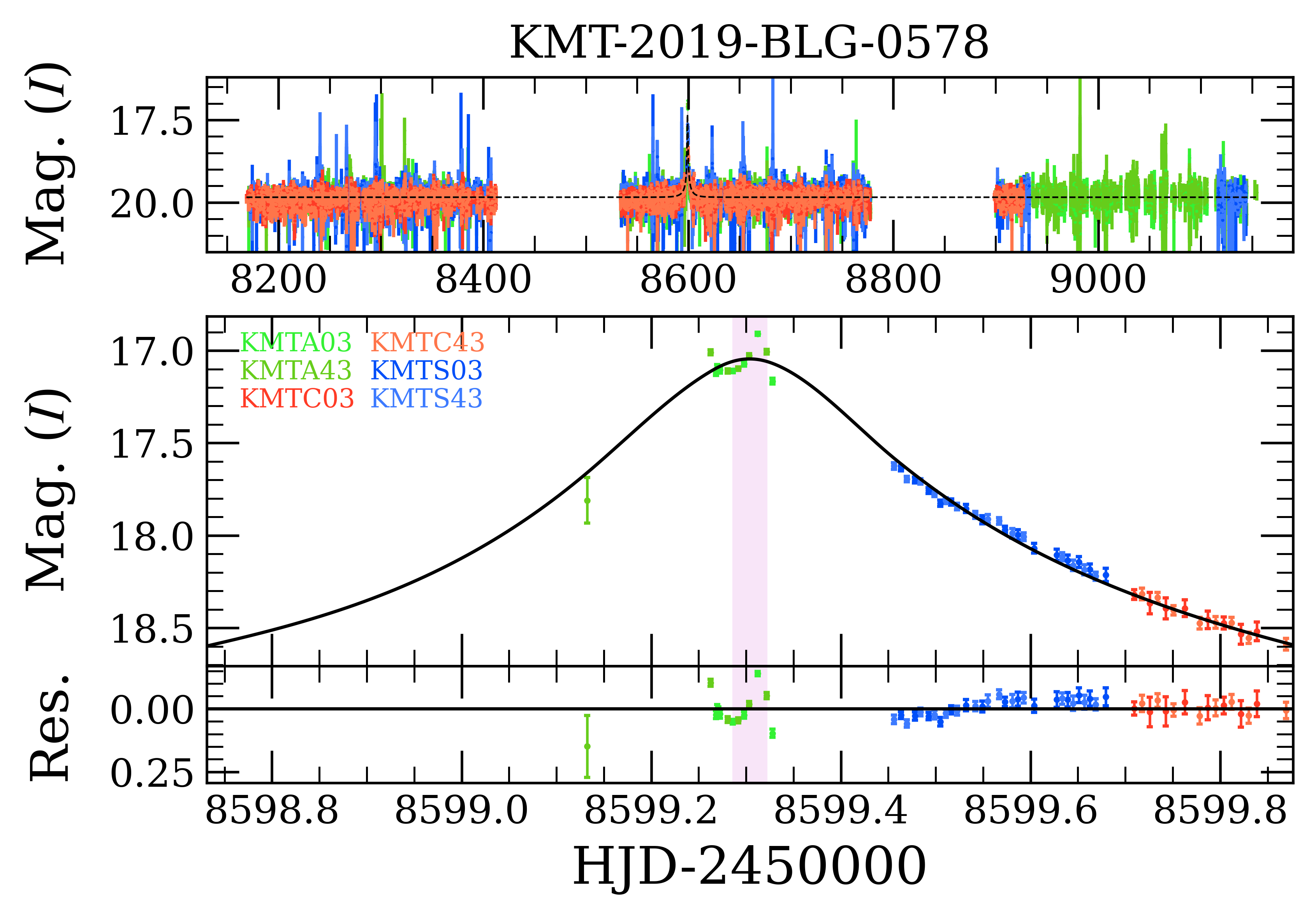}
    \\ \vspace{10pt}
    \includegraphics[width=0.9\columnwidth]{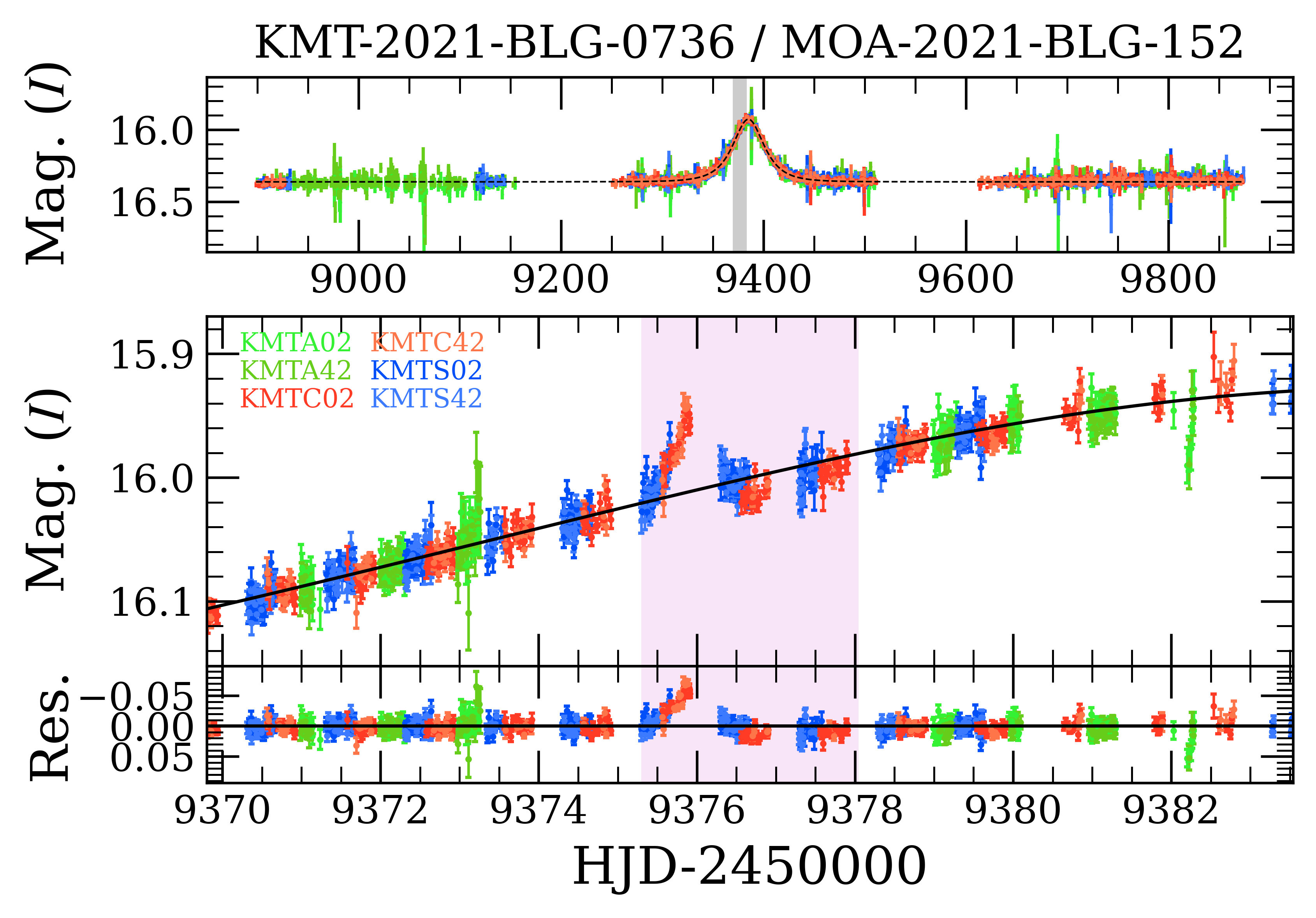}
    \includegraphics[width=0.9\columnwidth]{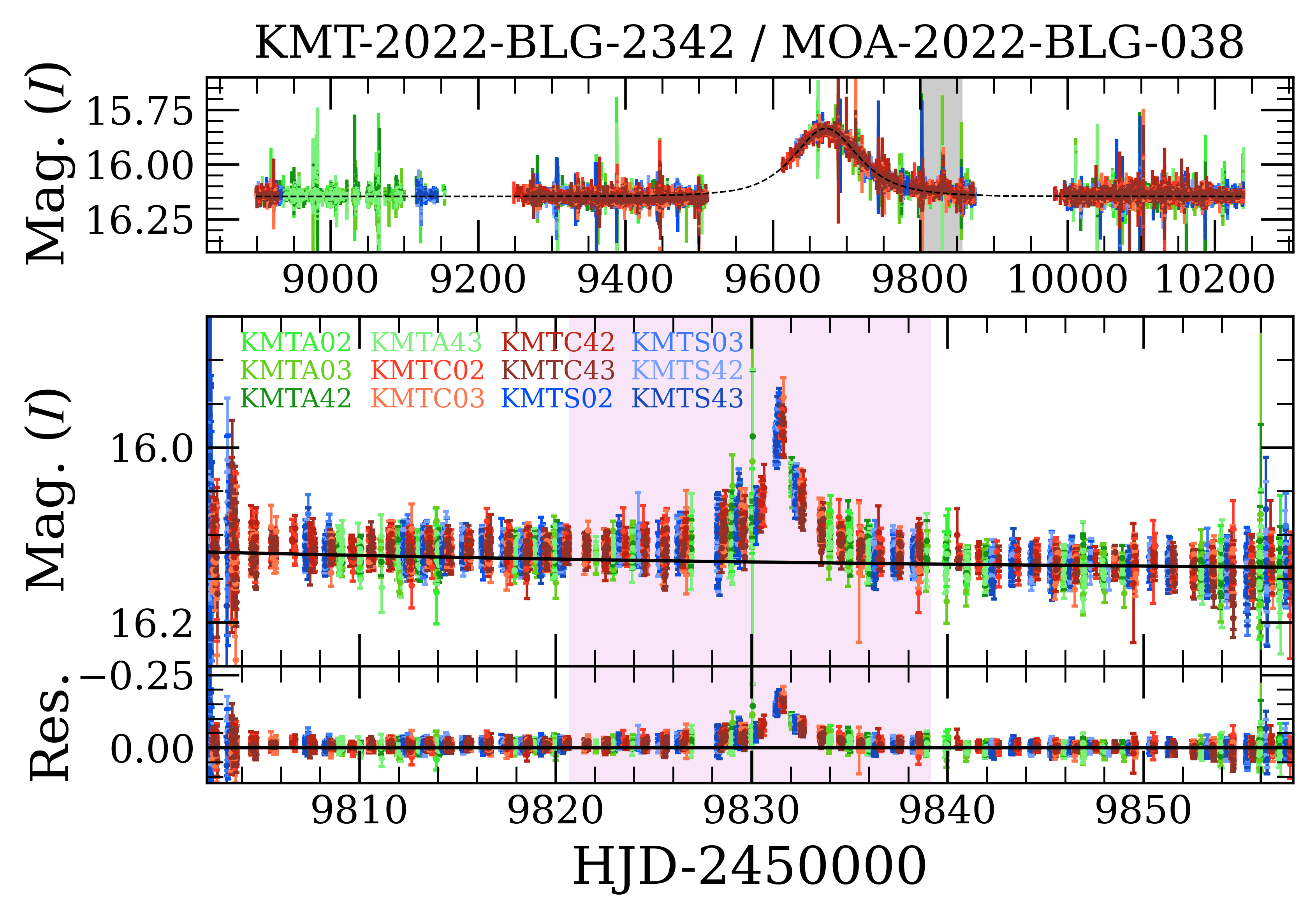}
    \caption{\HL{Same as Fig. \ref{fig:anomlc_noplanet} but for events with newly identified planet-like signals.}}
    \label{fig:anomlc_newcandidate}
\end{figure*}

    The other 4 events are considered new candidate planetary events, they are \KMT{18}{0831}, \KMT{19}{0578}, \KMT{21}{0736}, and \KMT{22}{2342}. The light curves and the reported signals for these four events are shown in Fig. \ref{fig:anomlc_newcandidate}. 
    
    Table \ref{tab:new_anom_info} lists the observational information of these four events. 
    \KMT{18}{0831} was first discovered by the OGLE Early Warning System \citep[EWS,][]{Udalski1994OGLEews,Udalski2003OGLEews} and named \OGLE{18}{0421}. It was later independently identified by the KMTNet post-season EventFinder system \citep{Kim2018a_EventFinder}. 
    \KMT{19}{0578} was only found by the KMTNet real-time AlertFinder \citep{Kim2018_KMTalgorithm} on April 29, 2019.
    \KMT{21}{0736} was first found by the KMTNet AlertFinder on May 10, 2021 and then independently discovered by the MOA collaboration on June 2, 2021 as \MOA{21}{152}. 
    \KMT{22}{2342} was first identified on February 24, 2022 by MOA collaboration and denoted as \MOA{22}{038}, and later independently found by the KMTNet post-season EventFinder system.
    
    Although the anomalies in these events were discovered using KMTNet data only, not all the events were initially identified by KMTNet. Hereafter, we use the names \HL{from the survey who first discovered them} for these events, i.e., \OGLE{18}{0421}, \KMT{19}{0578}, \KMT{21}{0736}, and \MOA{22}{038}.
    In the following Section \ref{sec:analysis}, we further investigate these four events to determine whether they are clear planets.



\section{Analysis of Planet Candidates}\label{sec:analysis}
In this section, we present the modeling details of the four planet-like anomalies identified in Section \ref{sec:eAF:apply}. Their light curves are shown in Fig.~\ref{fig:anomlc_newcandidate}.

The anomaly signals can be categorized into three groups based on their morphologies. 
The event in the first group is \KMT{19}{0578}. Its anomaly shows an ``M'' shape over the peak. This feature can only be produced by a second object in the lens system. Therefore, only binary-source single-lens (2L1S) models need to be explored for this event. 
The second group consists of \OGLE{18}{0421}. The anomaly signal shows a slight brightening followed by a dimming compared to a PSPL model. 
Such subtle asymmetries can potentially arise from microlensing parallax, xallarap effects, or the presence of a secondary lens.
However, the short duration ($\sim3$ days) of the anomaly makes the parallax explanation (due to Earth's orbital motion) unlikely.
Additionally, the anomaly timescale is significantly shorter than the overall event timescale (∼15 days). If caused by the source star's orbital motion (xallarap effect), the remaining light curve should also show periodic features \citep[e.g.,][]{Han_Gould_1997_1L2Sxarallap}.
We attempted to model the anomaly with both parallax and xallarap effects, but these models either provided poor fits to the light curve or resulted in unphysical parameters.
Therefore, only 2L1S models are needed to be explored for this event. 
The last group is \KMT{21}{0736} and \MOA{22}{038}, both of which have an additional peak separate from the main peak.
These features can be explained by either a cusp approach in the 2L1S scenario \citep{MaoPaczyski1991,Gould1992}, or the presence of a fainter, secondary source passing closer to the lens \citep[1L2S model,][]{Gaudi1998_1L2S}.
Therefore, both 2L1S and 1L2S models should be investigated for these two events.

While the detailed analysis is tailored to each event, a general procedure is followed.
In Section \ref{sec:ana:pre}, we first describe the general modelling methods. Then, the results for each individual event are presented in Sections \ref{sec:ana:kb180831}-\ref{sec:ana:kb222342}.



\subsection{Preamble}\label{sec:ana:pre}
\subsubsection{Light Curve Modelling}
    As mentioned above, we only explore 2L1S and 1L2S models in detail for these candidate events. 

    Standard 2L1S models require seven parameters to describe the magnification as a function of time, $A(t)$. The first three parameters, $(t_0,u_0,\tE)$, are identical to those used in the PSPL model. The difference is that $t_0$ and $u_0$ are defined relative to the magnification center of the two lenses\footnote{When $s\leq1$, the magnification center is the mass center. When $s>1$, the magnification center is located at $\frac{q}{1+q}(s-\frac{1}{s})$ from the primary star.}. 
    The three following parameters are $(s,q,\alpha)$: $s$ and $q$ represent the projected separation (in units of the angular Einstein radius) and the mass ratio of the two lenses, and $\alpha$ is the direction of source-lens relative proper motion. 
    The last parameter is $\rho$, the angular size of the source in units of the angular Einstein radius. 
    In addition to the magnification, two flux parameters, $(f_{{\rm S},i}, f_{{\rm B},i})$ represent the source and blend flux are needed for each dataset $i$. The total flux as a function of time is then
    \begin{equation}
        f_{i}(t) = f_{{\rm S},i} A(t) + f_{{\rm B},i}.
    \end{equation}
    We employ the \texttt{VBBinaryLensing} \citep{Bozza2010_VBBL,Bozza2018_VBBL} package to calculate the magnification for the 2L1S models.

    Finding all possible solutions for a light curve with such a large number of parameters, especially considering the highly non-linear behavior of $(s,q,\alpha)$, is not a trivial task.
    Therefore, we adopt a grid search approach followed by Markov-chain Monte-Carlo (MCMC) sampling to explore the parameter space and recover all possible solutions.
    The grid search starts with an initial set of parameters.
    By default, we sample 49 equally spaced values in $-1.2\leq\log s\leq1.2$, 61 equally spaced values in $-6.0\leq\log q\leq0.0$, 10 equally spaced values in $-3.4\leq\log\rho\leq-0.7$, and 16 equally spaced values in $0^{\circ}\leq\alpha<360^{\circ}$ as the initial parameters. 
    Then, for each set of initial parameters, we fix $(\log s,\log q,\log\rho)$ and allow all other parameters to vary. We use the MCMC sampler \texttt{emcee} \citep{emcee} to search for local minima of the $\chi^2$ function. 
    After the grid search, one or more local minima might be identified. The corresponding parameter sets represent candidate models that can potentially describe the light curve. 
    For each candidate model, we perform another round of MCMC sampling allowing all parameters to be free.
    This refines the parameter estimates and provides their uncertainties. We then compare the goodness of fit for the models and decide which one(s) to adopt.

    Although for simple cases like \KMT{21}{0736} and \MOA{22}{038}, 2L1S solutions can be found using analytical estimates \citep[e.g.,][]{Ryu2022_KMTMassProduction1}, we consistently follow the above grid search approach for all events to ensure that all possible 2L1S solutions are explored.

    The light curve for a 1L2S event is essentially the superposition of two individual 1L1S light curves. 
    Following \citet{MB12486_Hwang2013_1L2S}, at least eight parameters are required to model this scenario.
    Among them, $(t_{0,1},u_{0,1},\rho_1)$ and $(t_{0,2},u_{0,2},\rho_2)$ are the approach time, impact parameter, and the source size of the two sources, respectively. The two sources' Einstein timescales $\tE$ are assumed to be the same. Finally, a flux ratio parameter 
    \begin{equation}
        q_{F} = \frac{f_{\rm S,2}}{f_{\rm S,1}}
    \end{equation}
    is needed. The total flux during the event is then
    \begin{equation}
        f_i(t) = f_{{\rm S},i} \left[ A_1(t) + q_{F}A_2(t) \right] + f_{{\rm B},i}, \quad f_{{\rm S},i}\equiv f_{{\rm S},1,i},
    \end{equation}
    where $A_1(t)$ and $A_2(t)$ represent the magnifications from the 1L1S models for the primary and secondary sources, respectively.
    The flux ratio could vary depending on the observational band. 
    Therefore, an independent flux ratio parameter $q_{F,\lambda}$ needs to be included for each band $\lambda$.

    The fitting process for 1L2S models is generally more straightforward due to the linear nature of the superposition. 
    Therefore we initially set $(t_{0,1},u_{0,1},\rho_1,\tE)$ as the best-fit 1L1S parameters of the primary event.
    Then, we choose a reasonable initial value for $t_{0,2}$ around the location of the secondary peak in the light curve. Proper initial values for $(u_{0,2},\rho_2,q_{F})$ can also be estimated regarding the magnification excess.
    Finally, we optimize all the parameters using MCMC sampling to minimize $\chi^2$ and obtain the final parameters of the 1L2S model.

\subsubsection{Source Properties}\label{sec:source}
    This section describes the method for determining the source star's de-reddened magnitude and color. 
    This information can be used to measure the source angular radius $\thetas$, and ultimately $\thetaE$ and the \HL{lens-source} relative proper motion $\murel$
    \begin{equation}
        \thetaE = \frac{\thetas}{\rho}, \quad \mu_{\rm rel}=\frac{\thetaE}{\tE}.
        \label{eq:thetae}
    \end{equation}
    These parameters provide information on the physical properties of both the source and lens systems.

    We begin by measuring the source's $I$ and $V$ magnitudes within a $3.3'\times3.3'$  square region centered on the source.
    \HLb{We measure the flux of all stars on both $I$- and $V$- band master reference image using pyDIA\footnote{MichaelDAlbrow/pyDIA: Initial Release on Github, doi:10.5281/zenodo.268049}, and then calibrate the magnitudes to the OGLE-III \citep{OGLEIII} system and plot them on a Color-Magnitude Diagram (CMD).}
    Additionally, the source magnitude $I_{\rm S}$ can be obtained from the light curve modelling process.
    For single-source events, the source color $(V-I)_{\rm S}$ remains constant throughout the event. Therefore, it can be measured by performing a linear regression on the time series data of the $I$ and $V$ light curves.
    For binary-source events, the source color is measured by modelling the independent flux parameters of the light curve.

    After that, we measure the centroid of Red Clump (RC) stars \citep{Yoo2004,Nataf2013} on the CMD. This centroid is then used to calculate the offset between the source(s) and the RC stars,
    \begin{equation}
        \Delta[(V-I),I] = [(V-I),I]_{\rm S} -  [(V-I),I]_{\rm RC},
    \end{equation}
    where $[(V-I),I]_{\rm RC}$ is the RC centroid. 
    By comparing this offset with the known de-reddened color and magnitude of the Galactic bulge RC stars $[(V-I),I]_{\rm RC,0}$ \citep{Bensby2013,Nataf2013}, we can determine the de-reddened color and magnitude of the source(s),
    \begin{equation}
        [(V-I),I]_{\rm S, 0} = [(V-I),I]_{\rm RC, 0}+\Delta[(V-I),I].
    \end{equation}
    Finally, using $[(V-I),I]_{\rm S, 0}$, we can estimate the source star's angular size $\thetas$ according to \citet{Adams2018}. 
    If the light curve modelling provides a measurement or constraint on $\rho$, we can convert it to a measurement or constraint on $\thetaE$ and $\mu_{\rm rel}$ using Eq. \ref{eq:thetae}.

\subsubsection{Lens Properties}\label{sec:lens}
    For the events that are confirmed as planetary events, the light curve modelling only provides relative parameters $(s,q)$ of the lens system, but not the physical distance and mass. 
    Determining the physical properties of the lens system requires at least two of the following three parameters: microlensing parallax $\piE$, angular Einstein radius $\thetaE$, and the brightness of the lens star \citep[see also][]{Zang2022_ob180799}. 
    For example, if both $\piE$ and $\thetaE$ are measured, the mass of the lens system can be uniquely determined using \citep{Gould2000_Formalism}
    \begin{equation}
        \Ml = \frac{\thetaE}{\kappa \piE}, \quad \kappa = \frac{4 G}{c^2 {\rm AU}}\simeq 8.144~{\rm mas}/M_{\odot},
    \end{equation}
    where $G$ is the gravitational constant and $c$ is the speed of light.

    However, it is uncommon to measure both effects in a single event. 
    Moreover, directly measuring the lens star's brightness requires resolving the lens and source stars. This requires a wait of $\gtrsim5-10~{\rm yr}$ for the current large telescopes.
    In cases where none or only one of these parameters is available, we can only infer the physical properties of the lens system using a Bayesian approach with a Galactic model prior.

    The Galactic model contains information about  the stellar density, mass function, and the velocity distribution of the Milky Way. 
    We adopt these distributions from ``Model C'' in \citet{Yang2021_GalacticModel}. 
    We simulate a large number of microlensing events based on the Galactic model. 
    Each simulated event is then assigned a weight that considers both the microlensing event rate and the likelihood function obtained from the light curve modelling.
    Specifically, the weight for the $i$-th simulated event is
    \begin{equation}
        w_{i} = \Gamma_{i}\times \mathcal{L}_{i}(\tE) \mathcal{L}_{i}(\thetaE)\mathcal{L}_{i}(\bm{\pi}_{\rm E}),
        \label{eq:weihgt_bayes}
    \end{equation}
    where $\Gamma_{i}\propto\theta_{{\rm E},i} \mu_{{\rm rel},i}$ is the microlensing event rate, with $\mu_{{\rm rel},i}$ being the relative proper motion. $\mathcal{L}(\tE)$, $\mathcal{L}(\thetaE)$, and $\mathcal{L}(\bm{\pi}_{\rm E})$ are the likelihood functions for the corresponding parameters derived from the light curve modelling. 
    If a particular observable is not measured in the real event, the corresponding likelihood function is set to be uniform.
    In addition, \citet{Gould2022_MASADA} pointed out that events with confirmed planetary signals might exhibit a different $\murel$ distribution compared to other events, potentially due to observational bias.
    To account for this, when $\rho$ is unmeasured, an additional term of $\mu_{{\rm rel},i}^{-1}$ is incorporated into the weights.

    After properly weighting all the simulated events, we can obtain the posterior distributions of the physical properties of the lens system (host mass, planet mass, and planet-to-host separation).

\subsection{\OGLE{18}{0421}/\KMT{18}{0831}}\label{sec:ana:kb180831}
    Here we describe the exploration of 2L1S models for \OGLE{18}{0421}. We note that this event was also discovered and observed by OGLE collaboration \citep{OGLEIV}. 
    \HL{Therefore, data from OGLE are included in the following process.
    The images from the OGLE survey were taken in the $I$ band and the data were reduced by the \citet{Wozniak2000_diapl} difference image pipeline first and then the updated pySIS \citep{pysis,Yang2024_pysis5_RAMP1}.}
    
    A preliminary analysis revealed significant long-term deviations in the light curve data for $t>8250$. This deviation could be intrinsic variation of the source or unknown systematic errors.
    As this deviation (timescale $\sim 1~{\rm yr}$) is not correlated with the $\sim3~{\rm d}$ anomaly, we exclude data points beyond $8150\leq t\leq8250$ to avoid contamination.

\begin{figure*}
    \centering
        \includegraphics[width=1.8\columnwidth]{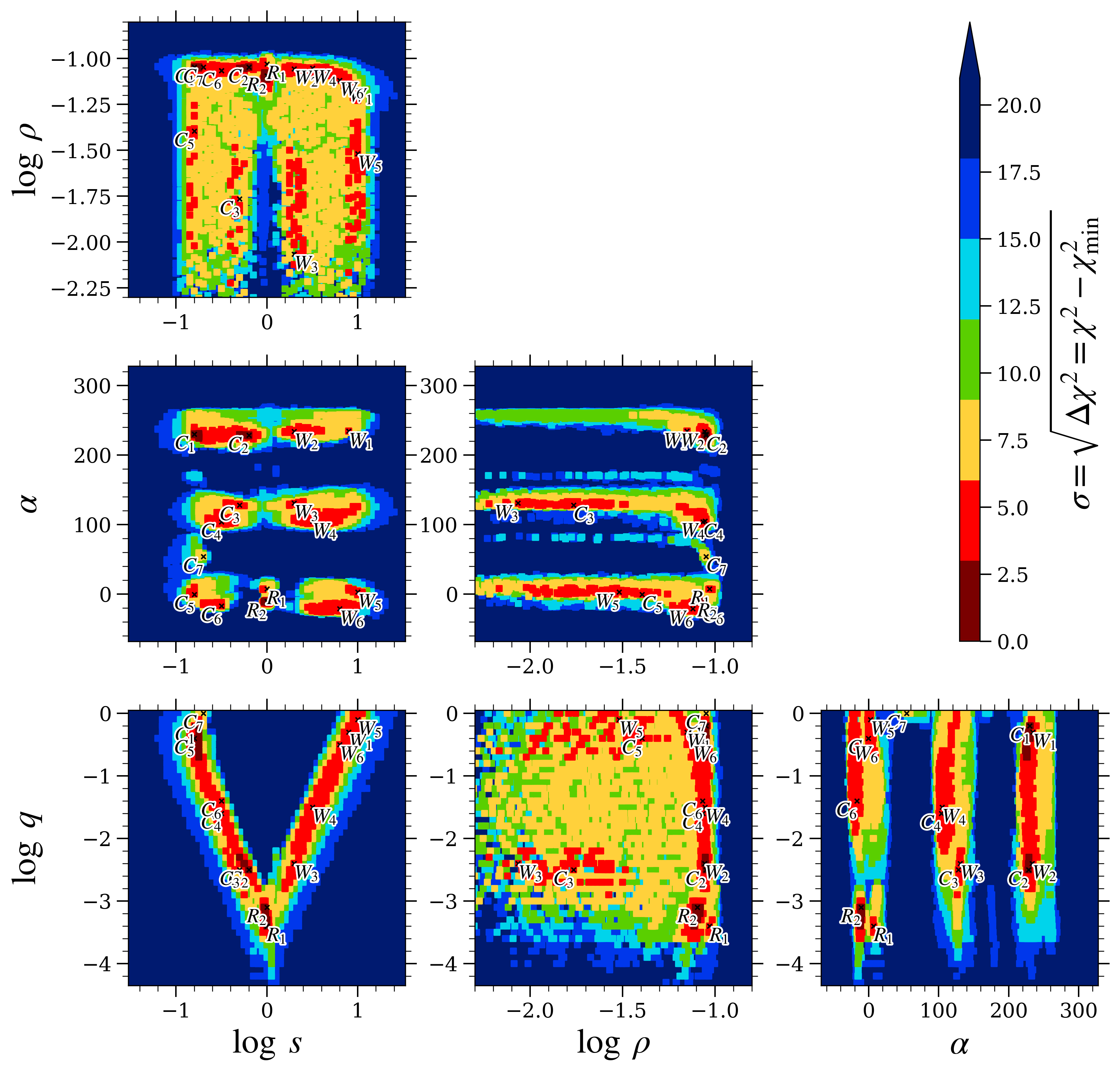}
    \caption{The 2L1S grid search $\chi^2$ distribution for \OGLE{18}{0421} in $(\log s, \log q, \alpha, \HL{\log\rho})$ space ($\alpha$ is units of degrees). The colors are coded by the $\Delta\chi^2$. \HL{The recognized local minima within 10$\sigma$ are labeled with their names.}}
    \label{fig:kb180831_grid}
\end{figure*}

    We first conduct a standard grid search \HL{as described in Section \ref{sec:ana:pre}. However, the local minima are not well covered, especially in $\log\rho$ space. Therefore, we adjust the default grid search to allow $\rho$ to vary in the optimization process. The adjusted grid search returns fifteen local minima within $\Delta\chi^2<100$, ($C_1$, $\cdots$, $C_7$, $R_1$, $R_2$, $W_1$, $\cdots$, $W_6$) as Fig. \ref{fig:kb180831_grid} shows.
    These initial models can be grouped into three categories regarding their mass ratios. Models with $\log q>-1$ ($C_1$, $C_5$, $C_7$, $W_1$, $W_5$, $W_6$) indicate stellar mass companions of the primary lens, models with $-2<\log q<-1$ ($C_4$, $C_6$, $W_4$) are likely brown dwarf companions, and $\log q<-2$ ($C_2$, $C_3$, $W_2$, $W_3$, $R_1$, $R_2$) indicate planetary mass companions. The subtle nature of the signal in the light curve results in strong degeneracy among stellar, brown dwarf, and planetary models.
    }

\begin{figure*}
    \centering
        \includegraphics[width=1.8\columnwidth]{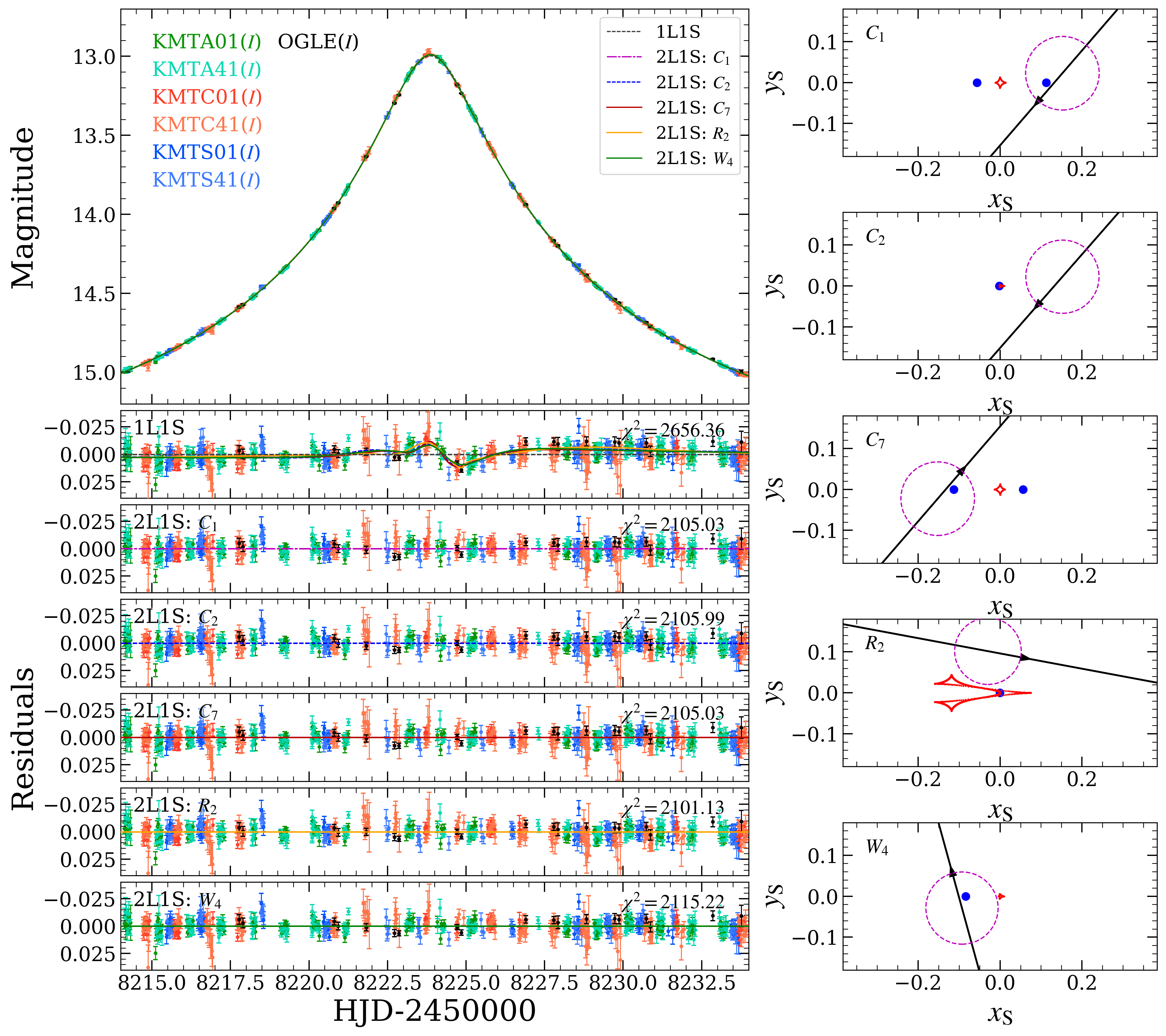}
    \caption{{\it Left:} The light curve, models, and residuals of \OGLE{18}{0421} around the anomalous region. The residuals of each model are shown in separate panels with their $\chi^2$. {\it Right:} Caustics and trajectories of 2L1S models for \OGLE{18}{0421}. The red curves are the caustics and the black lines with arrows are the trajectories of the source-to-lens motion. The dashed circles mark the finite source sizes $\rho$.} 
    \label{fig:kb180831_lc_cau}
\end{figure*}

    We then perform further optimization around each of the \HL{fifteen} initial local minima. 
    \HL{Table \ref{tab:kb180831_params} lists the optimized parameters of the remaining five solutions with $\Delta\chi^2<15$. 
    Model $R_2$ emerges as the best fit but only with a $\Delta\chi^2$ of approximately $(3.8, 4.0, 4.7, 14.1)$ compared to the other models $(C_1, C_7, C_2, W_4)$.
    }
    Fig. \ref{fig:kb180831_lc_cau} shows the light curves and models around the anomaly region for these five solutions \HL{in the left panels}.
    The corresponding caustic geometries and source-to-lens trajectories are shown in \HL{the right panels of} Fig. \ref{fig:kb180831_lc_cau}. 
    As shown in Table \ref{tab:kb180831_params} and Fig. \ref{fig:kb180831_lc_cau}, all three possible scenarios, stellar companion model ($C_1$, $C_7$), brown dwarf companion model ($W_4$), and planetary companion models ($C_2$, $R_2$) can describe the anomalous light curve well. 
    
    Notably, all \HL{five} degenerate solutions share consistent constraints on source brightness and $\rho$ constraints. 
    This implies that even future follow-up observations measuring the lens brightness and $\thetaE$ may not be sufficient to break the degeneracy.
    Therefore, we conclude that \OGLE{18}{0421} is an ambiguous event and do not investigate it further. 

\subsection{KMT-2019-BLG-0578}\label{sec:ana:kb190578}
    The event \KMT{19}{0578}, according to previous discussions, also needs only 2L1S modelling. 
    A standard grid search recovers only two local minima, $C$ and $W$ for this event. Unlike the previous event \OGLE{18}{0421}, the presence of strong caustic-crossing features in \KMT{19}{0578} limited the number of potential solutions.
    The two minima have similar values for \HLc{$\log q=-2.4$} and $\alpha\sim258^\circ$, but have opposite $\log s = \pm0.1$. 
    This again reflects the well-known ``close-wide degeneracy'' in microlensing \citep{Griest1998}.
    We then perform MCMC optimization around both local minima.
    The results are listed in Table \ref{tab:kb190578_params}. 
    
\begin{figure*}
    \centering
    \includegraphics[width=1.8\columnwidth]{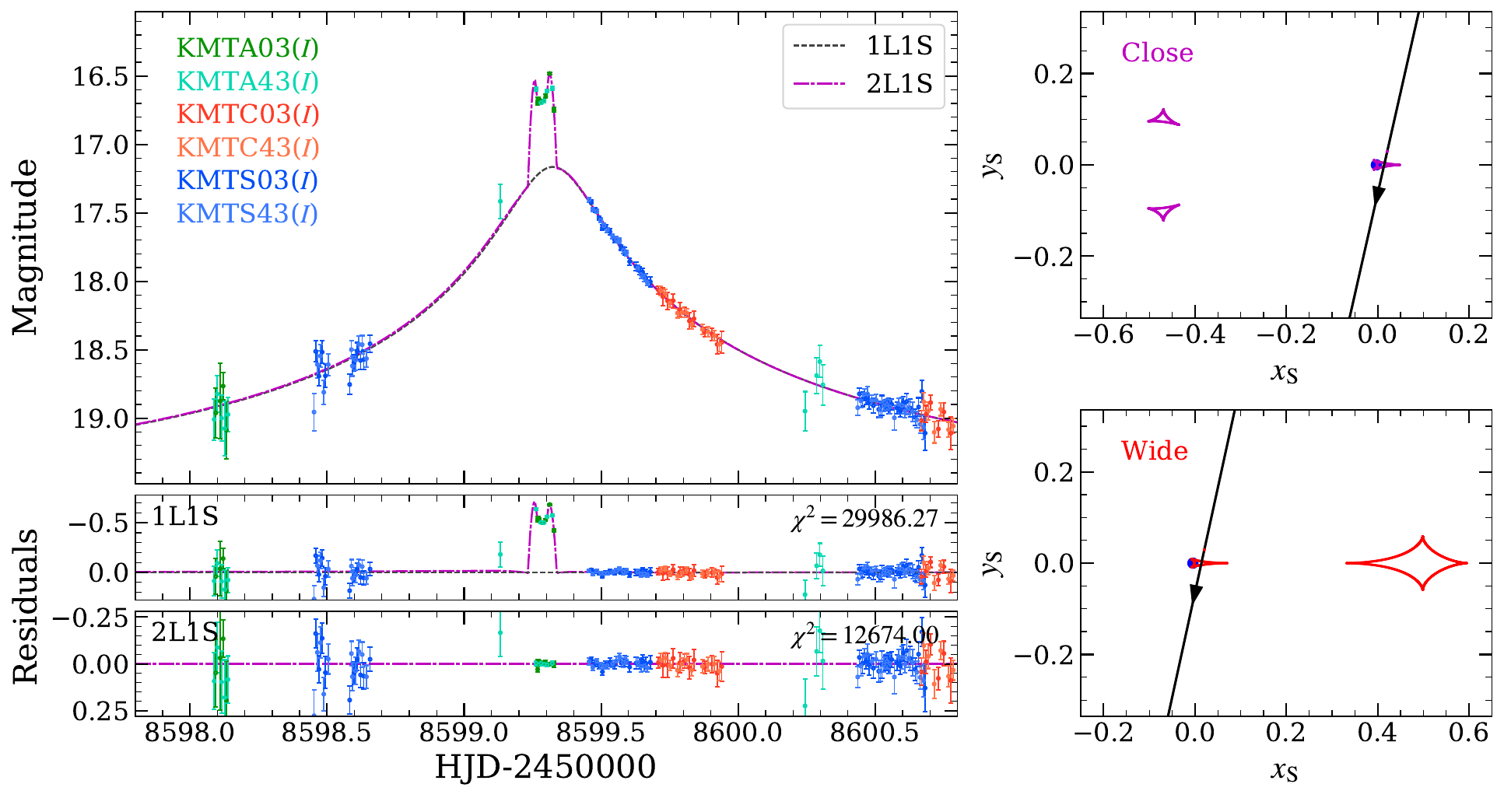}
    \caption{{\it Left:} The light curve, models, and residuals of \KMT{19}{0578} around the anomalous region. The residuals of each model are shown in separate panels and their $\chi^2$ are presented. The two 2L1S models are visually identical, thus only one ($C$) is presented. {\it Right:} Caustics and trajectories of 2L1S models for \KMT{19}{0578}. The curves are the caustics and the lines with arrows are the trajectories of the source-to-lens motion.}
    \label{fig:kb190578_lc_cau}
\end{figure*}

    The two models are almost identical and differ only by $\Delta\chi^2\sim0.02$. 
    \HL{The left panels of} Fig. \ref{fig:kb190578_lc_cau}. present the light curves along with the models. Because the two 2L1S models do not have visual differences, only the model $C$ is presented. 
    \HL{The right panels of} Fig. \ref{fig:kb190578_lc_cau} show the caustic structures and the lens-source motion trajectories of the two models. As expected, the central caustics of the two models are practically identical.

    Despite the degeneracy, both models consistently suggest the presence of a planet within the lens system, with a mass ratio of $q\sim4\times10^{-3}$.
    Therefore, the event \KMT{19}{0578} is a clear planetary event. 
    Because the models provide a measurement of $\rho$, we proceed to estimate $\thetaE$ using the method outlined in Section \ref{sec:source}

\begin{figure*}
    \centering
    \includegraphics[width=0.9\columnwidth]{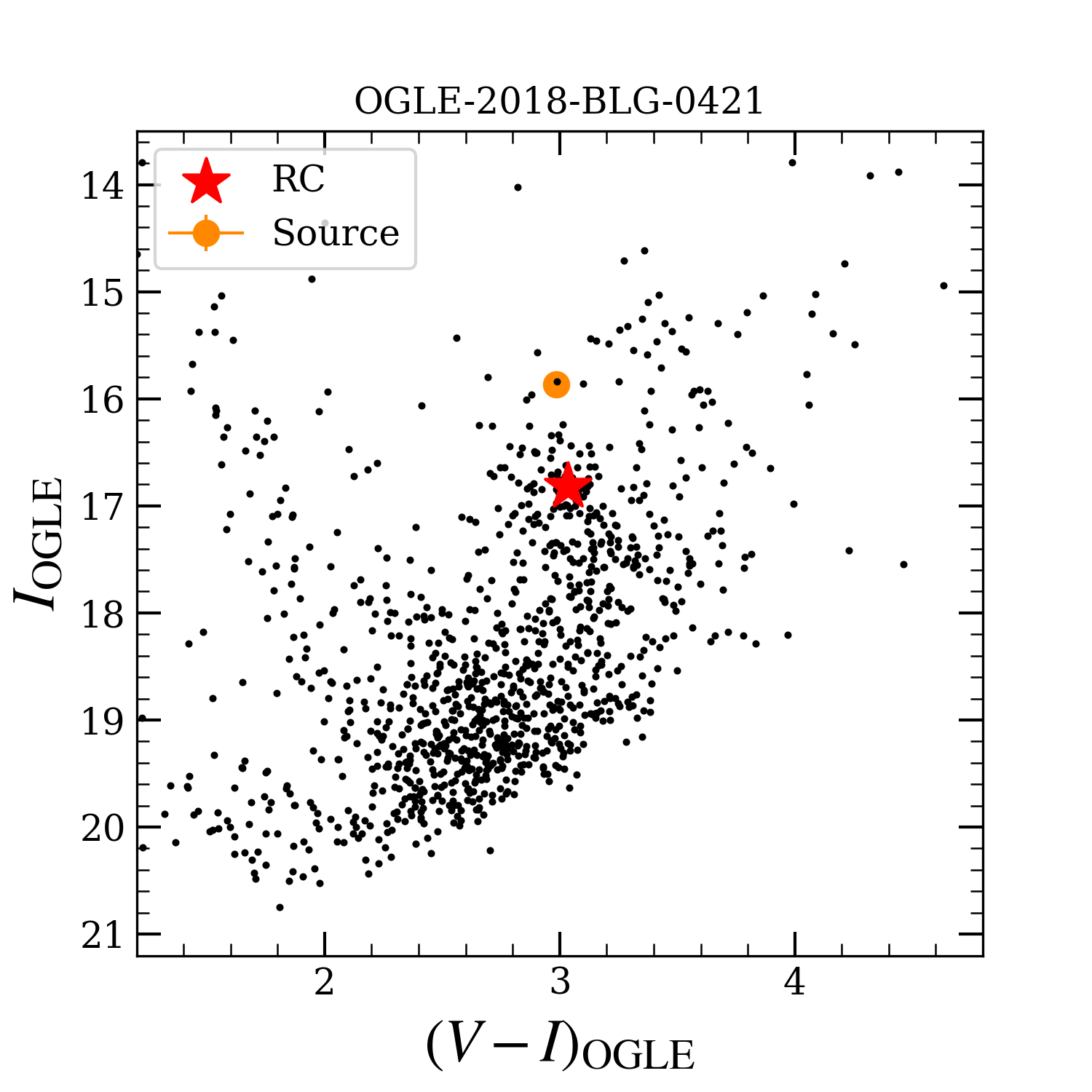}
    \includegraphics[width=0.9\columnwidth]{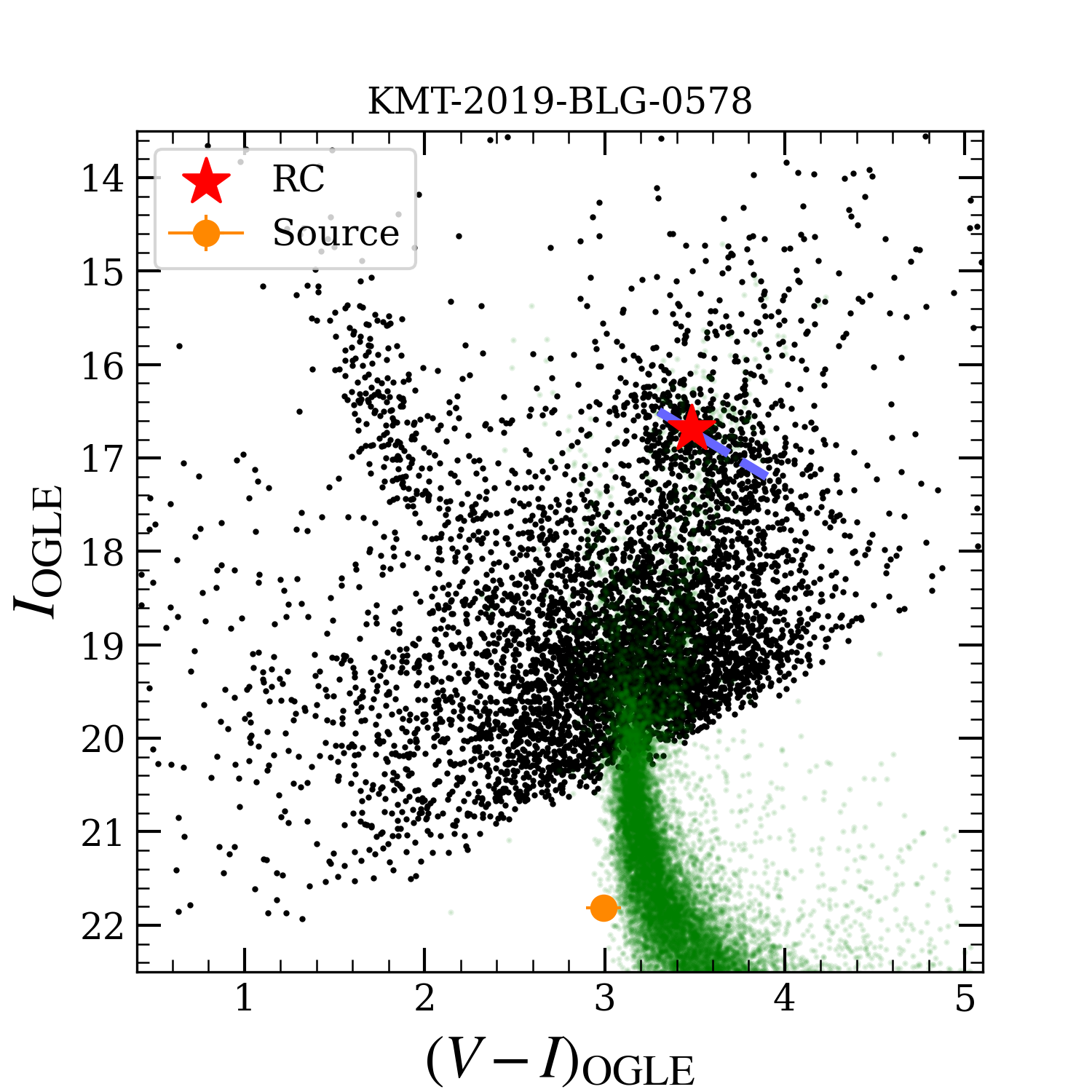}
    \\
    \includegraphics[width=0.9\columnwidth]{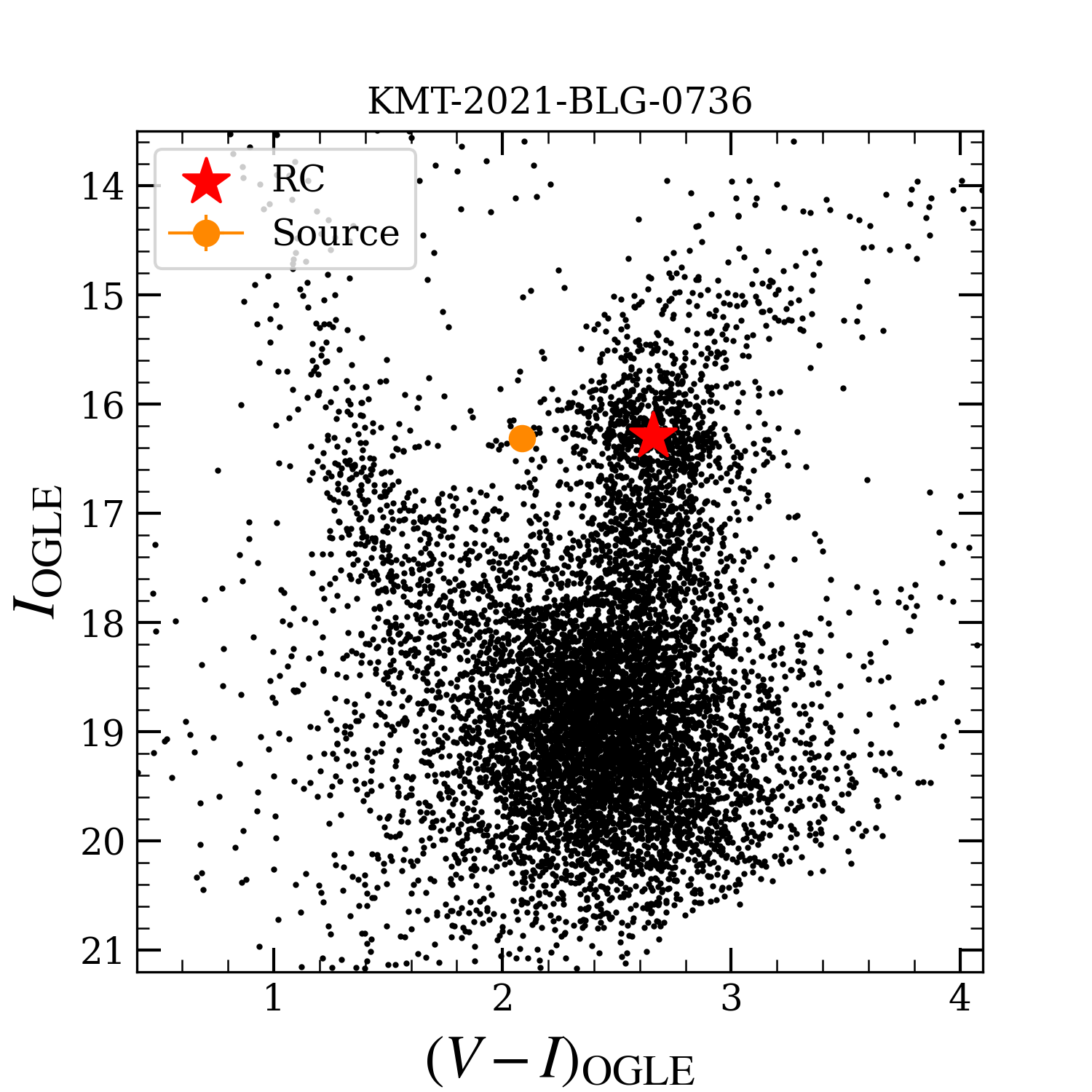}
    \includegraphics[width=0.9\columnwidth]{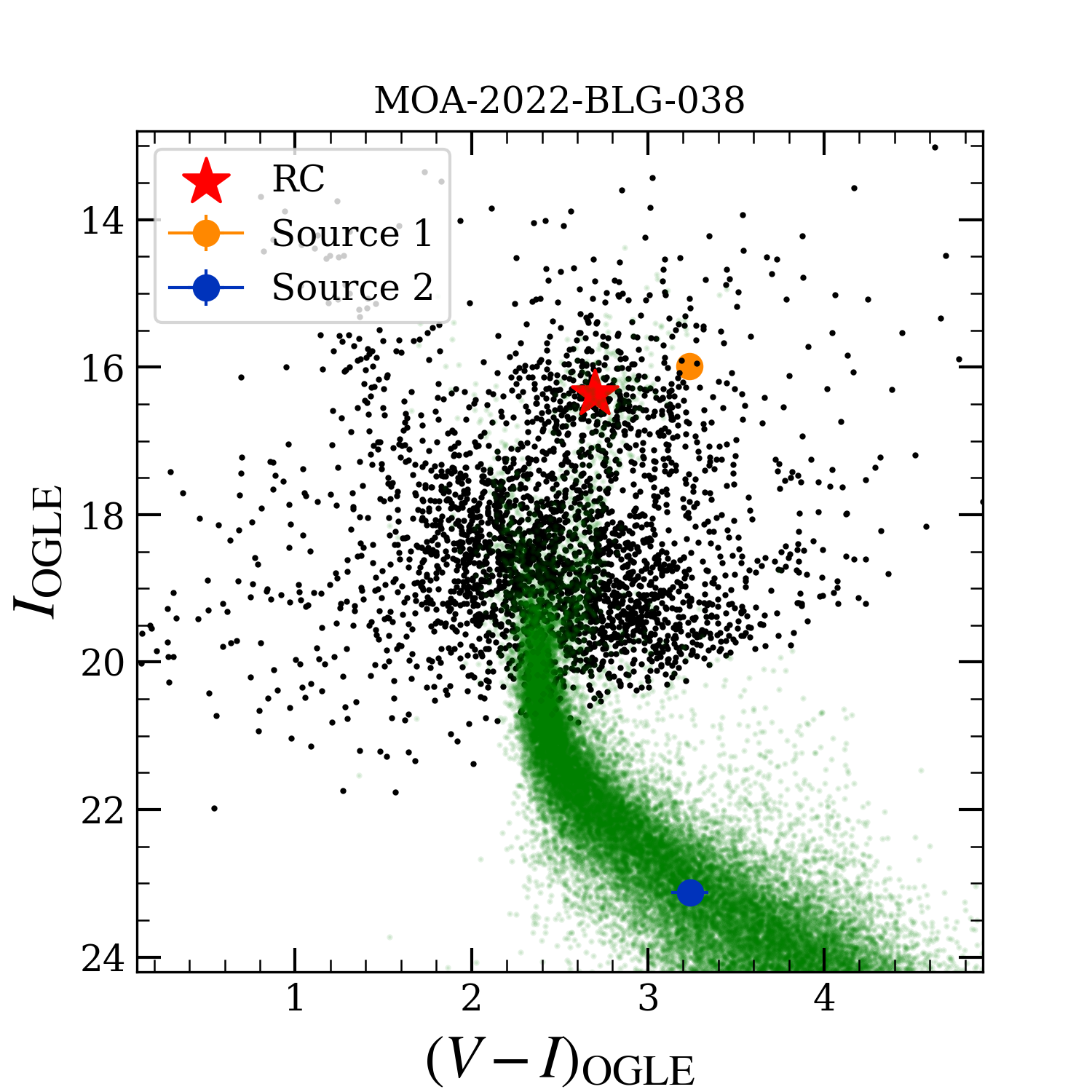}
    \caption{Color-Magnitude Diagrams (CMDs) for the four new candidate planetary events. The black dots are the field stars within $2'\times2'$ of each lensed source. The green dots stars are from HST observations toward Baade's window \citep{HSTCMD} and have been offset to match the RC centroid. Colors and magnitudes are calibrated to OGLE-III \citep{OGLEIII}. }
    \label{fig:cmd}
\end{figure*}

    We utilize the KMTA03 reference images to create the CMD and calibrate it to the OGLE-III catalog. The CMD is shown in Fig. \ref{fig:cmd}. 
    \HLc{The source star of this event is marked in the figure, it is offset to the bulge main sequence population.
    However, the red clump's CMD of this field seems elongated (the dashed blue line as a hint), which indicates that the field has a significant extinction variation. The source star could be a member of the low-extinction main-sequence population. 
    Nevertheless, no matter which population the source is in, the determined source angular radius $\thetas$ values is consistent within $\sim1\sigma$. Therefore, we still use the overall red clump centroid to continue the analysis.}
    Table \ref{tab:kb190578_source} summarizes the measured values, including the RC centroid, the source color and magnitude, and their de-reddened values. 
    The intrinsic RC centroid are calculated based on a linear interpretation of the values presented in \citet{Nataf2013}. Finally, we estimate the source size to be \HLc{$\thetas=0.361\pm0.040~{\rm \mu as}$} \citep{Adams2018}, and consequently \HLc{$\thetaE=0.213\pm0.029~{\rm mas}$} and \HLc{$\murel=8.8\pm1.3~{\rm mas/yr}$}.

    This event has a measurement of $\thetaE$, but the short timescale ($<10~{\rm d}$) prevents the constraints on the microlensing parallax. 
    We estimate the physical properties following the Bayesian approach mentioned in Section \ref{sec:lens}. We simulate $10^7$ microlensing events and assign weights based on Eq. \ref{eq:weihgt_bayes}. The median and $\pm 1\sigma$ confidence intervals of the resulting posterior distribution are presented in Table \ref{tab:kb190578_bayes}. 
    Based on the analysis, the planet is likely a giant planet with a mass of $\sim1.2\Mjup$ ($\Mjup$ denotes Jupiter mass). It is orbiting its \HLc{$\sim 0.28\Msun$} host at a projected distance of either \HLc{$\sim 1.2~$AU or $\sim 2.0~$AU}.

    In conclusion, although \KMT{19}{0578} has two degenerate models, both of them suggest a clear \HLc{$q\approx4\times10^{-3}$} planet in the lens system. \HL{The signal is not visible in the online KMTNet data.} 
    The discovery of this planet relies on the new photometric data reduced by the updated pipeline. 
    Using Galactic models, we estimate the planet to be a \HLc{$\sim1.2\Mjup$} planet orbiting its \HLc{$\sim0.28\Msun$} M dwarf host at a projected distance of about either \HLc{$1.2$ or $2.0~$AU}.

\subsection{KMT-2021-BLG-0736}\label{sec:ana:kb210736}
    Both 1L2S and 2L1S models can potentially describe the anomaly of \KMT{21}{0736}. Therefore, here we explore both of them. 
    
    Data from MOA are included in the modelling process.
    The images from the MOA survey were mainly taken in the MOA-$Red$ band, which is approximately the sum of the standard Cousins $R$ and $I$ bands. The MOA data were reduced by the \citet{Bond2001} difference image pipeline. 
    
\begin{figure*}
    \centering
        \includegraphics[width=1.8\columnwidth]{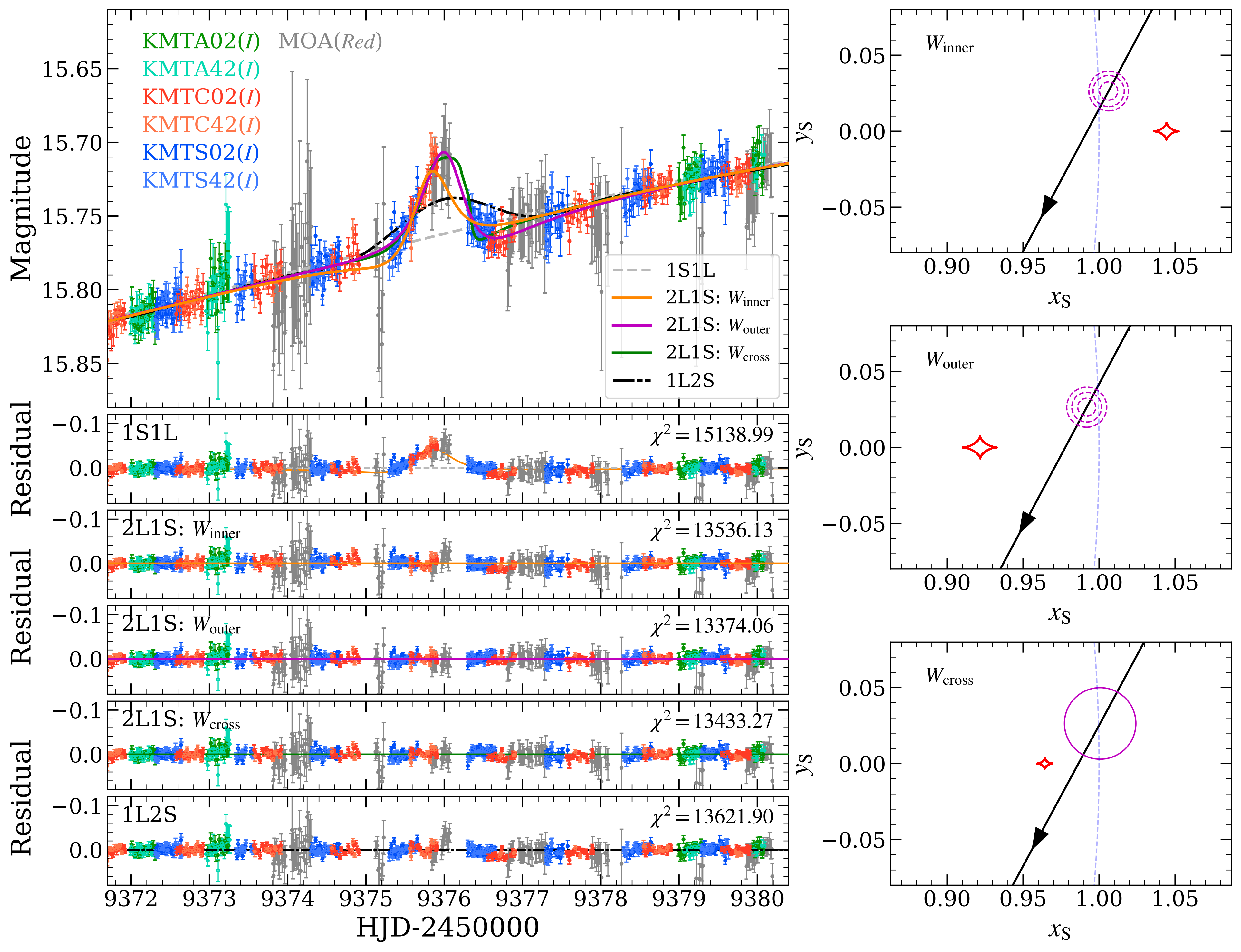}
    \caption{{\it Left:} The light curve, models, and residuals of \KMT{21}{0736} around the anomalous region. The residuals of each model are shown in separate panels and their $\chi^2$ are presented. {\it Right:} Caustics and trajectories of 2L1S models for \KMT{21}{0736}. The curves are the (planetary) caustics and the lines with arrows are the trajectories of the source-to-lens motion. The magenta dashed circles mark the $(1,2,3) \sigma$ upper limits of $\rho$ and the solid circle (bottom right) marks the measured values of $\rho$. The blue (nearly vertical) lines represent the Einstein radius.} 
    \label{fig:kb210736_lc_cau}
\end{figure*}

    \HL{ First, for 2L1S models, a standard grid search returns three solutions, $W_{\rm inner}$ with $(\log s, \log q)\sim(0.22,-4.2)$, $W_{\rm outer}$ with $(\log s, \log q)\sim(0.19,-4.0)$, and $W_{\rm cross}$ with $(\log s, \log q)\sim(0.20,-4.6)$. 
    We further optimized these solutions using a MCMC. The resulting parameters are summarized in Table \ref{tab:kb210736_params}. 
    The light curves and models around the anomaly region are presented in Fig. \ref{fig:kb210736_lc_cau}.}
    
    \HL{As illustrated in Fig. \ref{fig:kb210736_lc_cau}, these three models represent the source approaching/crossing the planetary caustics from different sides. 
    The degeneracy between $W_{\rm inner}$ and $W_{\rm outer}$ is known as the ``inner-outer'' degeneracy \citep{Gaudi1997_InnerOuterDenegeracy,KemingZhang2021_OffsetDegeneracy,Ryu2022_KMTMassProduction1}. }
    However, in this case, the $W_{\rm inner}$ model is significantly disfavored by $\Delta\chi^2\sim162$ and can be ruled out. 
    This is because $W_{\rm inner}$ and $W_{\rm outer}$ predict contrasting deviations in the light curve before and after the planetary peak. $W_{\rm inner}$ predicts a lower flux compared to the 1L1S model before the peak and a higher flux afterward, while $W_{\rm outer}$ suggests the opposite trend.
    The difference is well resolved thanks to the complete coverage of the observational data. 
    \HL{In addition, Model $W_{\rm cross}$ is a caustic-crossing solution, where the large source covers the entire planetary caustic. It predicts a shorter ``dip'' after the planetary peak, thus is disfavored by $\Delta\chi^2\sim59$. This model can also be ruled out.}
    Consequently, $W_{\rm outer}$ emerges as the only remaining 2L1S model for \KMT{21}{0736}.

    We then try 1L2S models for this event. The initial parameters for the first source were set to the best-fit values from the 1L1S model. The second source was initialized with  $(t_{0,2},u_{0,2},\rho_2)=(9376,0.01,0.01)$ and $q_{F,I}=q_{F,Red}=0.001$. We then performed an MCMC search with all parameters free.
    The final optimized parameters are listed in Table \ref{tab:kb210736_params}, and the corresponding model is plotted in Fig. \ref{fig:kb210736_lc_cau}. While the 1L2S model has two additional parameters compared to the 2L1S models, the best-fit solution yields a significantly poorer fit to the light curve.
    The figure also demonstrates the inability of the 1L2S model to capture the signal. Therefore, the 1L2S interpretation is excluded.
    
    In conclusion, the 2L1S $W_{\rm outer}$ is the only model of the light curve, indicating the presence of a planet in the lens system with a mass ratio of $q\approx1.06\times10^{-4}$.
    
    We then follow the methodologies in Sections \ref{sec:source}-\ref{sec:lens} to estimate the source and lens properties.
    The Color-Magnitude Diagram (CMD) of stars surrounding the source is constructed and displayed in Fig. \ref{fig:cmd}. The source of \KMT{21}{0736} is a red giant that is bluer than the Red Clump. 
    The measured de-reddened source color and magnitudes are \HLc{$(V-I,I)_{S,0}=(0.489,14.450)$}, leading to an estimated angular source size of \HLc{$\thetas=3.69\pm0.35~{\rm \mu as}$}. 
    Table \ref{tab:kb210736_source} provides the detailed measurements.
    
    Although the light curve model does not directly measure $\rho$, it provides constraints on its value. Based on the 3$\sigma$ upper limit of $\rho$, we find \HLc{$\thetaE>0.27~{\rm mas}$ and $\murel>5.0~{\rm mas/yr}$}.
    The above constraint on $\thetaE$ is considered in the Bayesian analysis. We simulate $10^8$ events using the Galactic model and weight them according to Eq. \ref{eq:weihgt_bayes}. 
    \HL{For bright source events, $Gaia$ observations could provide additional information. 
    We checked the source proper motion reported by $Gaia$ DR3 \citep{GaiaDR3_2023_summary}, however, the source star has a renormalized unit weight error (RUWE) of 1.63, which indicates a problematic astrometric solution. 
    We therefore measure the proper motion distribution for the red clump stars within a radius of 2$^{\prime}$ from the source, which has a mean value of $(\mu_{l},\mu_{b})=(-6.075,0.055)~{\rm mas/yr}$ and a covariance matrix of $C_{\mu_{l},\mu_{b}}=((12.053,-1.230),(-1.230,11.114))$. 
    The distribution is incorporated into the simulated events as an additional prior of the source star's proper motion.}
    The resulting posterior distributions of the physical parameters are presented in Table \ref{tab:kb210736_bayes}. 
    The estimated mass of the planet is \HLc{$21~\Mearth$}, and it orbits its \HLc{$\sim0.6\Msun$} host star at a distance of around \HLc{$4.0~$AU}.


    This is another new clear planetary event from the systematic search using new photometric data.

\subsection{\MOA{22}{038}}\label{sec:ana:kb222342}
    Event \MOA{22}{038} also needs both 2L1S and 1L2S modelling. 
    We note that this event is also located in MOA fields and was first discovered by MOA collaboration \citep{Bond2001,Sumi2003_MOAobs}. 
    \HL{Therefore, data from MOA are included in the following analysis.}
    
\begin{figure*}
    \centering
        \includegraphics[width=1.8\columnwidth]{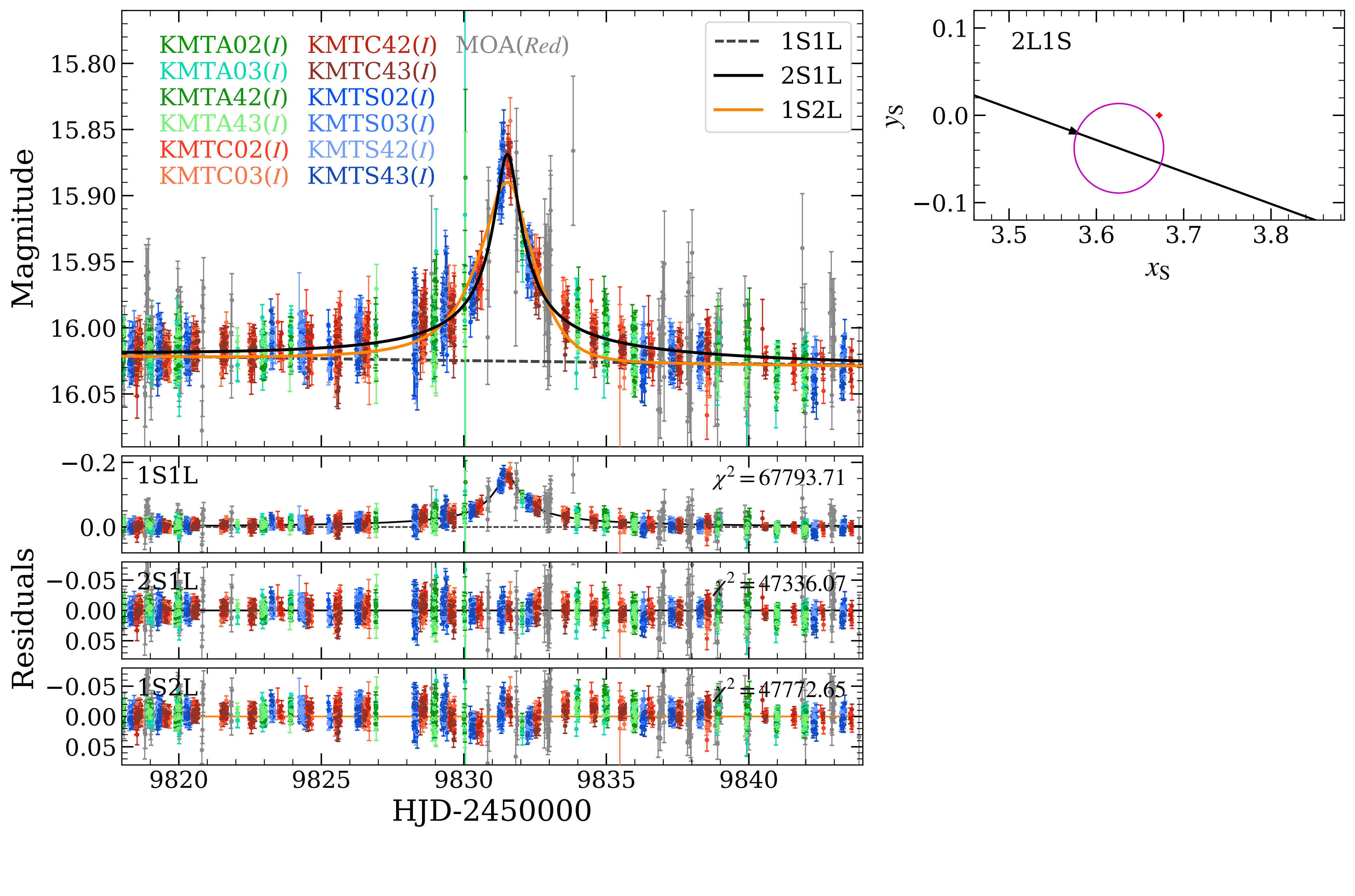}
    \caption{{\it Left:} The light curves, models, and residuals for \MOA{22}{038} around the anomalous region. The residuals of each model are shown in separate panels and their $\chi^2$ are presented. {\it Reft:} Caustics and trajectories of 2L1S models for \MOA{22}{038}. The curves are the (planetary) caustics and the lines with arrows are the trajectories of the source-to-lens motion. The magenta circle marks the source size $\rho$.} 
    \label{fig:kb222342_lc_cau}
\end{figure*}

    We first check 2L1S models. The signal is similar to \KMT{21}{0736}, thus we also expect \HL{a group of ``inner-outer-cross'' solutions.} 
    However, after detailed modelling, we find the expected solutions merge into a single one within the MCMC chains. The parameters of this unique solution are summarized in Table \ref{tab:kb222342_params}. 
    The reason is that the anomaly signal lasts relatively long and has relatively high magnification, which favors the scenario where the source entirely crosses a planetary caustic rather than merely approaching a cusp \HL{(see the right panel of Fig. \ref{fig:kb222342_lc_cau})}. Such a large $\rho$ smooths out the distinction between the ``inner'', ``outer'' \HL{and ``cross''} models, leading to the single solution. 
    The light curves around the anomaly, \HL{along with the model, residuals, and the corresponding caustic geometry} are shown in Fig. \ref{fig:kb222342_lc_cau}. 
    However, the presence of systematic residuals suggests that the 2L1S model might not be the best model for the event.

    We then investigate the 1L2S scenario. The initial parameters for the second source are set to $(t_{0,2},u_{0,2},\rho_2)=(9831.5,0.01,0.01)$, $q_{F,I}=q_{F,Red}=0.001$. The remaining parameters related to the first source are initialized with the values obtained from the 1L1S model. MCMC optimization is then performed with all parameters free. The resulting optimized model parameters and their uncertainties are listed in Table \ref{tab:kb222342_params}. Fig. \ref{fig:kb222342_lc_cau} also presents the model and residuals for this scenario. 

    When comparing the 1L2S model to the 2L1S model in Fig. \ref{fig:kb222342_lc_cau}, a clear improvement in the fit can be observed.
    The $\chi^2$ values also support this observation, with the 2L1S model being significantly disfavored by $\Delta\chi^2>400$. Based on this comparison, we exclude the 2L1S interpretation for this event.

    Despite favoring the 1L2S model, we perform an additional self-consistency check. 
    The event is located in the highest cadence fields of KMTNet, and the anomaly lasts about 5 days. Therefore, KMTNet provides many $V-$band observations during the anomaly.
    We include $V-$band data from KMTNet CTIO to measure the color information. The last row in Table \ref{tab:kb222342_params} lists the results obtained using all $I-$band data and KMTC $V-$band data. 
    The flux ratios $q_{F,I}$ and $q_{F,V}$ allow us to calculate the magnitude and color difference between the two sources as following,
    \begin{align}
        \Delta I_{\rm S} &= I_{\rm S,2} - I_{\rm S,1} = -2.5\log q_{F,I}, \\
        \Delta (V-I)_{\rm S} &= (V-I)_{\rm S,2}-(V-I)_{\rm S,1} =  -2.5\log\frac{q_{F,V}}{q_{F,I}}.
    \end{align}
    The positions of both sources are plotted on the CMD in Fig. \ref{fig:cmd}. 
    The first source is a red giant and the second source is a typical bulge main-sequence star. Therefore, the 1L2S model is reasonable, there is no evidence to contradict it.

    In summary, the analysis of the anomaly signal in event \MOA{22}{038} suggests the presence of a companion star to the the lensed source. The binary source system consists of a red giant and a main-sequence star.
    There is no indication that the lens system cannot be treated as a point lens. The newly discovered candidate anomaly is not a planetary signal.
    

\section{Conclusion and Discussion}\label{sec:summary}

In this work, we update the photometric pipeline based on \citet{Yang2024_pysis5_RAMP1} and form an automatic ``auto-TLC'' pipeline. 
We define a giant source event sample based on the KMTNet database, including a total of 352 events. We then apply the auto-TLC pipeline to these events to produce high-quality photometric data. In this sample, the ``auto-TLC'' photometry \HL{is successfully produced for 340 events.} 
We then develop an anomaly search algorithm, \texttt{eAF}, and use it to identify potential planetary signals within the light curves of these events. 
We recovered 2 previously known planetary signals and 16 previously known stellar binary signals in the sample, and find 4 new planet-like anomalies. 
The events with new candidate planetary anomalies are \OGLE{18}{0421} (\KMT{18}{0831}), \KMT{19}{0578}, \KMT{21}{0736}, and \MOA{22}{2342} (\KMT{22}{2342}).

\HL{Subsequent detailed modelling of the detected anomalies revealed that the nature of the anomalies in \HLb{\KMT{19}{0578} and \KMT{21}{0736} are clear planetary signals.}}
The planet in event \KMT{19}{0578} has a mass-ratio of $q\sim4\times10^{-3}$. It is likely to be a \HLc{$\sim1.2$} Jovian mass planet orbiting the M dwarf host at a distance of \HLc{$\sim1.2$ or $\sim2.0~$AU}, depending on which one of the degenerate solutions is correct.
The M/K dwarf lens star in \KMT{21}{0736} hosts a $q\sim1\times10^{-4}$ (or \HLc{$21\Mearth$}) Neptune-mass planet. The (projected) orbital distance is \HLc{$\sim4.0~$AU}.
\HLb{The other two events \OGLE{18}{0421} and \MOA{22}{2342} remains ambiguous or suggests non-planetary interpretations.}

\begin{figure*}
    \centering
        \includegraphics[width=1.0\columnwidth]{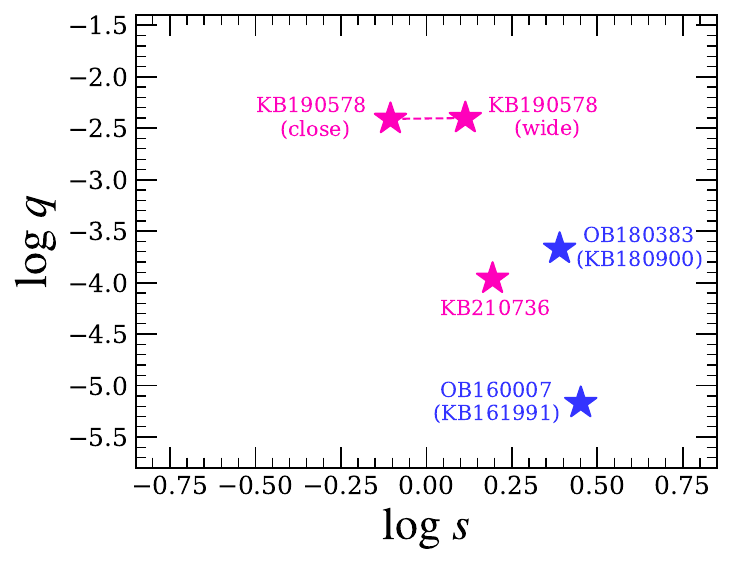}
    \caption{The $(\log s, \log q)$ distribution of the newly discovered (pink) and recovered (blue) clear planets in the sample.} 
    \label{fig:clear_planet}
\end{figure*}

By systematically re-analyzing the 340 giant-source events with improved photometry, we successfully identified two new planetary systems. This approach has led to the discovery of previously missed planetary signals, effectively doubling the number of confirmed planets within the analyzed sample. 
Figure \ref{fig:clear_planet} shows all the four planets in this sample in $(\log q, \log s)$ space.

To rigorously answer the question of how much planet-detection efficiency is improved, a comprehensive evaluation of sensitivity is needed. 
Fortunately, the \texttt{eAF} anomaly search algorithm developed here offers the potential to systematically calculate the sensitivity of the sample to planet detections. 
The detailed sensitivity calculations as well as the corresponding statistical analysis will be presented in a future paper.

\acknowledgments
\HLc{H.Yang acknowledges support by the China Postdoctoral Science Foundation (No. 2024M762938).}
H.Yang, J.Zhang, W.Zang, Q.Qian, and S.Mao acknowledge support by the National Natural Science Foundation of China (Grant No. 12133005). 
J.C.Y. and I.-G.S. acknowledge support from U.S. NSF Grant No. AST-2108414. 
This research has made use of the KMTNet system operated by the Korea Astronomy and Space Science Institute (KASI) at three host sites of CTIO in Chile, SAAO in South Africa, and SSO in Australia. Data transfer from the host site to KASI was supported by the Korea Research Environment Open NETwork (KREONET). This research was supported by KASI under the R\&D program (project No. 2024-1-832-01) supervised by the Ministry of Science and ICT.
\HLc{The MOA project is supported by JSPS KAKENHI Grant Number 
JP24253004, JP26247023,JP16H06287 and JP22H00153.}
The authors acknowledge the Tsinghua Astrophysics High-Performance Computing platform at Tsinghua University for providing computational and data storage resources that have contributed to the research results reported within this paper. 

{\it Software:} NumPy \citep{numpy:2020}, SciPy \citep{scipy:2020}, astropy \citep{astropy:2013,astropy:2018,astropy:2022}, \href{https://doi.org/10.5281/zenodo.268049}{pyDIA}, PyAstronomy \citep{PyAstronomy}, VBBinaryLensing \citep{Bozza2010_VBBL,Bozza2018_VBBL}

\begin{table*}
    \centering
    \caption{Observational information of the new candidate planetary events}
    \renewcommand{\arraystretch}{1.2}
    \label{tab:new_anom_info}
    \begin{tabular}{lrrrr}
    \hline
    KMTNet name & KB180831 & KB190578$^*$ & KB210736$^*$ & KB222342 \\ 
    \hline
    Other name  & OB180421$^*$ & $-$      &  MB21152 &  MB22038$^*$ \\
    R.A.~(J2000) &    17:51:58.73 &    17:57:56.32 &    17:53:17.42 &    17:58:04.53 \\
    Dec.~(J2000) & $-$31:51:58:10 & $-$27:47:41.50 & $-$29:42:44.71 & $-$28:17:16.30 \\
    $l$ & $-$1.806$^{\circ}$ &    2.364$^{\circ}$ &    0.194$^{\circ}$ &    1.952$^{\circ}$ \\
    $b$ & $-$2.684$^{\circ}$ & $-$1.748$^{\circ}$ & $-$1.833$^{\circ}$ & $-$2.021$^{\circ}$ \\
    KMTNet field & 01,41 & 03,43 & 02,42 & 02,03,42,43 \\ \hline
    \end{tabular}
    \begin{tablenotes}
      \item NOTE. KB180831 is the abbreviation of \KMT{18}{0831}, OB180421 is the abbreviation of \OGLE{18}{0421}, and MB21152 is the abbreviation of \MOA{21}{152}, and so on. The \HL{official name (based on first discovery)} is marked by ``$^*$''.
    \end{tablenotes}
\end{table*}

\begin{table*}
    \centering
    \caption{\HL{The 2L1S parameters for \OGLE{18}{0421}}}
    \renewcommand{\arraystretch}{1.2}
    \label{tab:kb180831_params}
    \begin{tabular}{lrrrrrr}
        \hline
        Model & $C_1$ & $C_2$ & $C_7$ & $\bm{R_2}$ & $W_4$ \\ 
        \hline
        $\chi^2/{\rm dof.}$       & $      2105.03/2103$ & $      2105.99/2103$ & $      2105.03/2103$ & $\bm{      2101.13/2103}$ & $      2115.22/2103$ \\
        $t_0$~(HJD$^\prime-$8223) & $   0.8585\pm0.0031$ & $8223.8812\pm0.0011$ & $8223.8607\pm0.0037$ & $\bm{8223.8925\pm0.0021}$ & $8223.8799\pm0.0011$ \\
        $u_{0}$                   & $   0.1003\pm0.0005$ & $   0.0997\pm0.0005$ & $   0.1001\pm0.0005$ & $\bm{   0.0945\pm0.0006}$ & $   0.0968\pm0.0010$ \\
        $t_{\rm E}$~(d)           & $  15.710 \pm0.051 $ & $  15.730 \pm0.040 $ & $  15.728 \pm0.050 $ & $\bm{  16.004 \pm0.043 }$ & $  16.257 \pm0.155 $ \\
        $\rho$                    & $   0.0898\pm0.0008$ & $   0.0893\pm0.0011$ & $   0.0897\pm0.0009$ & $\bm{   0.0810\pm0.0016}$ & $   0.0876\pm0.0013$ \\
        $\alpha$~($^\circ$)       & $ 229.1   \pm1.9   $ & $ 228.9   \pm1.5   $ & $  48.3   \pm1.9   $ & $\bm{ -10.64  \pm0.59  }$ & $ 105.6   \pm1.5   $ \\
        $\log s$                  & $  -0.773 \pm0.012 $ & $  -0.244 \pm0.048 $ & $  -0.771 \pm0.015 $ & $\bm{  -0.0252\pm0.0027}$ & $   0.540 \pm0.061 $ \\
        $\log q$                  & $  -0.34  \pm0.11  $ & $  -2.40  \pm0.13  $ & $   0.44  \pm0.16  $ & $\bm{  -3.108 \pm0.036 }$ & $  -1.43  \pm0.18  $ \\
        $s$                       & $   0.1686\pm0.0047$ & $   0.574 \pm0.058 $ & $   0.170 \pm0.006 $ & $\bm{   0.9437\pm0.0058}$ & $   3.50  \pm0.50  $ \\
        $q$                       & $   0.47  \pm0.10  $ & $   0.0042\pm0.0023$ & $   3.0   \pm1.3   $ & $\bm{ 0.00078\pm0.00006}$ & $   0.041 \pm0.017 $ \\
        $f_{\rm S,KMTC01}$        & $   8.667 \pm0.039 $ & $   8.666 \pm0.033 $ & $   8.654 \pm0.039 $ & $\bm{   8.388 \pm0.036 }$ & $   8.515 \pm0.045 $ \\
        $I_{\rm S}$               & $  15.8744\pm0.0052$ & $  15.8745\pm0.0045$ & $  15.8760\pm0.0052$ & $\bm{  15.9099\pm0.0049}$ & $  15.8936\pm0.0061$ \\
        \hline
    \end{tabular}
    \begin{tablenotes}
        \centering
        \item{* HJD$^\prime$=HJD-2450000. Model parameters and their $1\sigma$ uncertainties are presented. For unmeasured parameters, the $3\sigma$ limit are provided. The magnitudes have been calibrated to OGLE-III. \HL{The best solution is marked in boldface.}}
    \end{tablenotes}
\end{table*}

\begin{table*}
    \centering
    \caption{The 2L1S parameters for \KMT{19}{0578}}
    \renewcommand{\arraystretch}{1.2}
    \label{tab:kb190578_params}
    \begin{tabular}{lrr}
        \hline
        Model              &      $C$ &      $W$  \\ 
        \hline
        $\chi^2/{\rm dof.}$ &    $  12674.00/12673$ &  $   12673.98/12673$  \\
        $t_0$~(HJD$^\prime$) & $8599.3102\pm0.0034$ & $8599.3096\pm0.0033$  \\
        $u_{0}$              & $   0.0137\pm0.0010$ & $   0.0137\pm0.0011$  \\
        $t_{\rm E}$~(d)      & $   8.84  \pm0.48  $ & $   8.83  \pm0.50  $  \\
        $\rho~(10^{-3})$     & $   1.67  \pm0.12  $ & $   1.69  \pm0.13  $  \\
        $\alpha$~($^\circ$)  & $ 257.4   \pm1.4   $ & $ 257.8   \pm1.4   $  \\
        $\log s$             & $  -0.105 \pm0.011 $ & $   0.113 \pm0.012 $  \\
        $\log q$             & $  -2.408 \pm0.063 $ & $  -2.401 \pm0.072 $  \\
        $s$                  & $   0.786 \pm0.019 $ & $   1.298 \pm0.036 $  \\
        $q~(10^{-4})$        & $  39.5   \pm5.8   $ & $  40.3   \pm6.7   $  \\
        $f_{\rm S,KMTC01}$   & $   0.0364\pm0.0026$ & $   0.0365\pm0.0027$  \\
        $I_{\rm S}$          & $  21.820 \pm0.074 $ & $  21.816 \pm0.070 $  \\
        \hline
    \end{tabular}
    \begin{tablenotes}
        \centering
        \item{* HJD$^\prime$=HJD-2450000. Model parameters and their $1\sigma$ uncertainty are presented. The magnitudes have been calibrated to OGLE-III.}
    \end{tablenotes}
\end{table*}

\begin{table*}
    \centering
    \caption{\HLc{Source properties and the derived microlensing parameters of \KMT{19}{0578}}}
    \renewcommand{\arraystretch}{1.2}
    \label{tab:kb190578_source}
    \begin{tabular}{lrr}
        \hline
        Parameter            &  Value  &  Uncertainty \\ 
        \hline
        $(V-I)_{\rm RC}$     &  3.484   & 0.021 \\
        $I_{\rm RC}$         & 16.689   & 0.036 \\
        $(V-I)_{\rm RC,0}$   &  1.060   & 0.030 \\
        $I_{\rm RC,0}$       & 14.365   & 0.040 \\ 
        \hline
        $(V-I)_{\rm S}$      &  2.994   & 0.095 \\
        $I_{\rm S}$          & 21.816   & 0.070 \\
        $(V-I)_{\rm S,0}$    &  0.570   & 0.102 \\
        $I_{\rm S,0}$        & 19.491   & 0.088 \\ 
        \hline
        $\thetas$~(${\rm \mu as}$) & 0.361 & 0.040 \\
        $\thetaE$~(${\rm mas}$)    & 0.213 & 0.029 \\
        $\murel$~(${\rm mas/yr}$)  & 8.9 &  1.3 \\
        \hline
    \end{tabular}
\end{table*}

\begin{table*}
    \centering
    \caption{\HLc{Physical properties from Bayesian analysis of planetary event \KMT{19}{0578}}}
    \renewcommand{\arraystretch}{1.25}
    \label{tab:kb190578_bayes}
    \begin{tabular}{lrr}
        \hline
        Model              & $C$ &  $W$  \\ 
        \hline
        $D_{\rm S}$~(kpc)  & $8.5^{+0.9}_{-0.7}$ & $8.5^{+0.9}_{-0.7}$ \\
        $D_{\rm L}$~(kpc)  & $7.3^{+0.7}_{-0.8}$ & $7.3^{+0.7}_{-0.8}$ \\
        $\murel$~(mas/yr)  & $8.8^{+1.3}_{-1.1}$ & $8.8^{+1.3}_{-1.1}$  \\
        $M_{\rm host}$~($M_\odot$)  & $0.28^{+0.28}_{-0.14}$ & $0.28^{+0.28}_{-0.14}$ \\
        $M_p$~($\Mjup$)  & $1.2^{+1.2}_{-0.6}$ & $1.2^{+1.2}_{-0.6}$ \\
        $a_{\perp}$~(AU)  & $1.2^{+0.2}_{-0.2}$ & $2.0^{+0.3}_{-0.3}$ \\
        \hline
    \end{tabular}
\end{table*}

\begin{table*}
    \centering
    \caption{The 1L2S and 2L1S parameters for \KMT{21}{0736}}
    \renewcommand{\arraystretch}{1.2}
    \label{tab:kb210736_params}
    \begin{tabular}{lrrrr}
        \hline
        Model                    & 1L2S & 2L1S: $W_{\rm inner}$ & \textbf{2L1S}:  $\bm{W_{\rm outer}}$ & \HL{2L1S: $W_{\rm cross}$}  \\ 
        \hline
        $\chi^2/{\rm dof.}$      &   $  13621.90/13372$ &   $  13536.13/13374$ & $\bm{    13374.06/13374}$ &   $  13433.27/13374$ \\
        $t_{0,1}$~(HJD$^\prime$) & $9385.284 \pm0.013 $ & $9385.198 \pm0.012 $ & $\bm{9385.196 \pm0.012 }$ & $9385.222 \pm0.012 $ \\
        $t_{0,2}$~(HJD$^\prime$) & $9375.844 \pm0.013 $ & $\cdots$ & $\cdots$ & $\cdots$ \\
        $u_{0,1}$                & $   0.815 \pm0.013 $ & $   0.871 \pm0.017 $ & $\bm{   0.863 \pm0.019 }$ & $   0.866\pm0.017$ \\
        $u_{0,2}$                & $  -0.0001\pm0.0011$ & $\cdots$ & $\cdots$ & $\cdots$ \\
        $t_{\rm E}$~(d)          & $  20.92  \pm0.21  $ & $  20.03  \pm0.26  $ & $\bm{  20.12  \pm0.29  }$ & $  20.09  \pm0.26  $ \\
        $\rho_{1}$               & $             <0.27$ & $           <0.012 $ & $\bm{           <0.013 }$ & $   0.0236\pm0.0007$ \\
        $\rho_{2}~(10^{-3})$     & $   7.06  \pm0.51  $ & $\cdots$ & $\cdots$ & $\cdots$ \\
        $q_{F,I}~(10^{-4})$      & $   2.71  \pm0.19  $ & $\cdots$ & $\cdots$ & $\cdots$ \\
        $q_{F,Red}~(10^{-4})$    & $   3.01  \pm0.46  $ & $\cdots$ & $\cdots$ & $\cdots$ \\
        $\alpha$~($^\circ$)      &             $\cdots$ & $ 241.88  \pm0.17  $ & $\bm{ 241.94  \pm0.18  }$ & $ 241.59  \pm0.16  $ \\
        $\log s$                 &             $\cdots$ & $   0.2164\pm0.0034$ & $\bm{   0.1940\pm0.0039}$ & $   0.2016\pm0.0034$ \\
        $\log q$                 &             $\cdots$ & $  -4.120 \pm0.025 $ & $\bm{  -3.972 \pm0.020 }$ & $  -4.613 \pm0.028 $ \\
        $s$                      &             $\cdots$ & $   1.646 \pm0.013 $ & $\bm{   1.563 \pm0.014 }$ & $   1.591 \pm0.012 $ \\
        $q~(10^{-4})$            &             $\cdots$ & $   0.760 \pm0.044 $ & $\bm{   1.068 \pm0.049 }$ & $   0.244 \pm0.016 $ \\
        $f_{\rm S,1,KMTC01}$     & $   4.73  \pm0.13  $ & $   5.40  \pm0.20  $ & $\bm{   5.33  \pm0.22  }$ & $   5.34  \pm0.20  $ \\
        \HLc{$I_{\rm S,1}$}      & $  16.419 \pm0.023 $ & $  16.293 \pm0.023 $ & $\bm{  16.313 \pm0.022 }$ & $  16.300 \pm0.025 $ \\
        \hline
    \end{tabular}
    \begin{tablenotes}
        \centering
        \item{* HJD$^\prime$=HJD-2450000. Model parameters and their $1\sigma$ uncertainty are presented. For unmeasured parameters, their $3\sigma$ limits are provided. The magnitudes have been calibrated to OGLE-III. The final selected model is highlighted in boldface.}
    \end{tablenotes}
\end{table*}

\begin{table*}
    \centering
    \caption{\HLc{Source properties and the derived microlensing parameters of \KMT{21}{0736}}}
    \renewcommand{\arraystretch}{1.2}
    \label{tab:kb210736_source}
    \begin{tabular}{lrr}
        \hline
        Parameter          &  Value  &  Uncertainty \\ 
        \hline
        $(V-I)_{\rm RC}$   &  2.657 & 0.008 \\
        $I_{\rm RC}$       & 16.297 & 0.022 \\
        $(V-I)_{\rm RC,0}$ &  1.060 & 0.030 \\
        $I_{\rm RC,0}$     & 14.434 & 0.040 \\ 
        \hline
        $(V-I)_{\rm S}$    &  2.086 & 0.030 \\
        $I_{\rm S}$        & 16.313 & 0.022 \\
        $(V-I)_{\rm S,0}$  &  0.489 & 0.043 \\
        $I_{\rm S,0}$      & 14.450 & 0.051 \\ 
        \hline
        $\thetas$~(${\rm \mu as}$) &   3.69  & 0.35 \\
        $\thetaE$~(${\rm mas}$)    & $>0.27$ &  $-$ \\
        $\murel$~(${\rm mas/yr}$)  &  $>5.0$ &  $-$ \\
        \hline
    \end{tabular}
\end{table*}

\begin{table*}
    \centering
    \caption{\HLc{Physical properties from Bayesian analysis of planetary event \KMT{21}{0736}}}
    \renewcommand{\arraystretch}{1.25}
    \label{tab:kb210736_bayes}
    \begin{tabular}{lr}
        \hline
        Model              & $W_{\rm outer}$  \\ 
        \hline
        $D_{\rm S}$~(kpc)  & \HLc{$8.9^{+0.7}_{-0.8}$} \\
        $D_{\rm L}$~(kpc)  & \HLc{$7.0^{+0.8}_{-1.5}$} \\
        $\murel$~(mas/yr)  &  \HLc{$6.9^{+1.8}_{-1.3}$}  \\
        $M_{\rm host}$~($M_\odot$)  & \HLc{$0.60^{+0.37}_{-0.29}$} \\
        $M_p$~($\Mearth$)  & \HLc{$21^{+13}_{-10}$} \\
        $a_{\perp}$~(AU)  & \HLc{$4.0^{+0.9}_{-0.8}$} \\
        \hline
    \end{tabular}
\end{table*}

\begin{table*}
    \centering
    \caption{The 1L2S and 2L1S parameters for \MOA{22}{038}/\KMT{22}{2342}}
    \renewcommand{\arraystretch}{1.2}
    \label{tab:kb222342_params}
    \begin{tabular}{lrrr}
        \hline
        Model                    & 2L1S & \textbf{1L2S} & \begin{tabular}[c]{@{}r@{}}1L2S\\ (+ KMTC $V$-band)\end{tabular}  \\ 
        \hline
        $\chi^2/{\rm dof.}$          & $    47772.65/47300$ & $\bm{    47336.15/47299}$ & $    48937.30/48891$ \\
        $t_{0,1}$~(HJD$^\prime$)     & $9671.674 \pm0.043 $ & $\bm{9671.816 \pm0.037 }$ & $9671.830 \pm0.044 $ \\
        $t_{0,2}$~(HJD$^\prime$)     &             $\cdots$ & $\bm{9831.5312\pm0.0092}$ & $9831.5314\pm0.0099$ \\
        $u_{0,1}$                    & $   1.236 \pm0.022 $ & $\bm{   1.0616\pm0.0009}$ & $   1.0615\pm0.0009$ \\
        $u_{0,2}$                    &             $\cdots$ & $\bm{   0.0087\pm0.0005}$ & $   0.0087\pm0.0004$ \\
        $t_{\rm E}$~(d)              & $  45.78  \pm0.55  $ & $\bm{  50.412 \pm0.056 }$ & $  50.406 \pm0.063 $ \\
        $\rho_{1}$                   & $   0.0528\pm0.0020$ &                  $\cdots$ &             $\cdots$ \\
        $\rho_{2}~(10^{-3})$         &             $\cdots$ & \HL{$\bm{        <9.9 }$} & \HL{$        <9.9 $} \\
        $q_{F,I}~(10^{-3})$          &             $\cdots$ & $\bm{   1.240 \pm0.016 }$ & $   1.245 \pm0.014 $ \\
        $q_{F,Red}~(10^{-3})$        &             $\cdots$ & $\bm{   1.300 \pm0.100 }$ & $   1.314 \pm0.097 $ \\
        $q_{F,V}~(10^{-3})$          &             $\cdots$ &                  $\cdots$ & $   1.325 \pm0.214 $ \\
        $\alpha$~($^\circ$)          & $ 339.76  \pm0.12  $ &                  $\cdots$ &             $\cdots$ \\
        $\log s$                     & $   0.5993\pm0.0049$ &                  $\cdots$ &             $\cdots$ \\
        $\log q$                     & $  -3.509 \pm0.014 $ &                  $\cdots$ &             $\cdots$ \\
        $s$                          & $   3.975 \pm0.045 $ &                  $\cdots$ &             $\cdots$ \\
        $q~(10^{-4})$                & $   3.10  \pm0.10  $ &                  $\cdots$ &             $\cdots$ \\
        $f_{{\rm S,1},I,{\rm KMTC01}}$ & $   9.18  \pm0.38  $ & $\bm{   6.561 \pm0.012 }$ & $   6.560 \pm0.012 $ \\
        $I_{\rm S,1}$                  & $  15.638 \pm0.025 $ & $\bm{  15.9906\pm0.0024}$ & $  15.9906\pm0.0026$ \\
        $f_{{\rm S,1},V,{\rm KMTC01}}$ &             $\cdots$ &                  $\cdots$ & $   0.5081\pm0.0010$ \\
        $V_{\rm S,1}$                  &             $\cdots$ &                  $\cdots$ & $  19.2261\pm0.0031$ \\
        \hline
    \end{tabular}
    \begin{tablenotes}
        \centering
        \item{* HJD$^\prime$=HJD-2450000. Model parameters and their $1\sigma$ uncertainty are presented. For unmeasured parameters, their $3\sigma$ limits are provided. No useful $\rho_1$ is measured in the 1L2S models. The magnitudes have been calibrated to OGLE-III. The final selected model is highlighted in boldface.}
    \end{tablenotes}
\end{table*}

\clearpage
\bibliography{Yang.bib}

\end{CJK*}
\end{document}